\documentclass[review,1p]{elsarticle}

\usepackage{dcolumn}
\usepackage{bm}
\usepackage{subfigure}

\usepackage[normalem]{ulem}

\usepackage{amsmath,amssymb,amsfonts}

\usepackage{blindtext}
\usepackage[capitalise]{cleveref}
\usepackage{cancel}

\newcommand\BibTeX{{\rmfamily B\kern-.05em \textsc{i\kern-.025em b}\kern-.08em
T\kern-.1667em\lower.7ex\hbox{E}\kern-.125emX}}

\newcommand{\bu}{\boldsymbol{u}}
\newcommand{\bv}{\boldsymbol{v}}

\newcommand{\ba}{\boldsymbol{a}}
\newcommand{\bc}{\boldsymbol{c}}
\newcommand{\bff}{\boldsymbol{f}}
\newcommand{\bz}{\boldsymbol{z}}
\newcommand{\bh}{\boldsymbol{h}}
\newcommand{\bs}{\boldsymbol{s}}

\newcommand{\bphi}{\boldsymbol{\phi}}
\newcommand{\bpsi}{\boldsymbol{\psi}}
\newcommand{\bPsi}{\boldsymbol{\Psi}}

\newcommand{\btau}{\boldsymbol{\theta}}
\newcommand{\bTheta}{\boldsymbol{\Theta}}

\newcommand{\bepsilon}{\boldsymbol{\epsilon}}

\usepackage{graphics}
\usepackage{color}
\usepackage{xcolor}
\definecolor{rev}{rgb}{0,0,0}

\usepackage{graphicx}
\usepackage{subfigure}
\usepackage{algorithm}
\usepackage{algpseudocode}
\algnewcommand\server{\item[\textbf{Server execution:}]}%
\algnewcommand\client{\item[\textbf{ClientUpdate($k,w$):}]}%

\usepackage{enumitem}
\setlist[itemize]{leftmargin=*}
\setlist[enumerate]{leftmargin=*}

\usepackage{comment}
\usepackage{lipsum}
\usepackage{tabularx}
\usepackage{multirow}
\usepackage{mathrsfs}
\usepackage{stackengine}

\usepackage{accents}

\usepackage{url,microtype}
\usepackage[onehalfspacing]{setspace}
\usepackage[thinc]{esdiff} 
\usepackage{amssymb}
\usepackage{amsmath}
\usepackage{mathtools}
\usepackage{bm}
\usepackage{xcolor}
\usepackage{setspace}
\usepackage{geometry}
\usepackage{tikz}
\usepackage{booktabs} 
\definecolor{green}{rgb}{0,0.6,0}

\usepackage{siunitx}
\usepackage[switch]{lineno}

\usepackage{placeins}

\usepackage{algorithm}
\usepackage{algpseudocode}

\journal{Elsevier}

\begin{document}

\begin{frontmatter}
\title{Forward Sensitivity Analysis and Mode Dependent Control for Closure Modeling of Galerkin Systems} 

\author[OSUaddress]{Shady E. Ahmed}
\ead{shady.ahmed@okstate.edu}

\author[OSUaddress]{Omer San\corref{mycorrespondingauthor}}
\cortext[mycorrespondingauthor]{Omer San}
\ead{osan@okstate.edu}



\address[OSUaddress]{School of Mechanical and Aerospace Engineering, Oklahoma State University}

\doublespacing
\begin{abstract}
Model reduction by projection-based approaches is often associated with losing some of the important features that contribute towards the dynamics of the retained scales. As a result, a mismatch occurs between the predicted trajectories of the original system and the truncated one. We put forth a framework to apply a continuous time control signal in the latent space of the reduced order model (ROM) to account for the effect of truncation. We set the control input using parameterized models by following energy transfer principles. Our methodology relies on observing the system behavior in the physical space and using the projection operator to restrict the feedback signal into the latent space. Then, we leverage the forward sensitivity method (FSM) to derive relationships between the feedback and the desired mode-dependent control. We \textcolor{rev}{test} the performance of the proposed approach using two test cases, corresponding to viscous Burgers and vortex merger problems at high Reynolds number. Results show that the ROM trajectory with the applied FSM control closely matches its target values in both the data-dense and data-sparse regimes.      
\end{abstract}

\begin{keyword}
Reduced order models \sep forward sensitivity \sep inverse problem  \sep latent control \sep outer-loop applications \sep sparse sensors
\end{keyword}
\end{frontmatter}

\section{Introduction} \label{sec:intro}
Engineers always tend to increase gains and reduce costs. For example, the \textcolor{rev}{airfoil} design of an airplane wing is optimized to increase lift, reduce drag, and enhance stability. The design optimization process involves multiple forward runs to simulate the system's response to different inputs, parameters, and operating conditions. This multi-query nature is often labeled as outer-loop applications while the individual forward simulations are known as inner-loop computations. For high dimensional systems (e.g., fluid flows), the wall-clock time for such computations becomes incompatible with desired turnaround times for design cycles as well as realtime control. This computational burden presents a roadblock to the routine use of simulation tools by industry. Therefore, lightweight surrogates are often sought to approximate the effective dynamics and reduce the computational \textcolor{rev}{costs} of inner-loop computations without compromising the integrity of the computational pipeline \cite{bergmann2009enablers,balajewicz2016minimal,kramer2017sparse,benosman2017learning,benosman2018model,peherstorfer2018survey,poveda2019hybrid,taira2020modal,huang2020balanced,riffaud2021dgdd,koc2022verifiability,papapicco2022neural,ivagnes2022towards}.

With the advent of data-driven tools and open-source software libraries, machine learning (ML) algorithms have been exploited to build computationally light emulators solely from data. The complex input-output relationships are learnt from precollected recordings of the system's dynamics during a compute-intensive process known as training. More recently, there has been an increasing interest in embedding existing knowledge to build hybrid physics informed ML frameworks \cite{san2021hybrid,kashinath2021physics,cuomo2022scientific,vlachas2022multiscale}, possibly by considering feature enhancement \cite{maulik2019subgrid}, using prediction from simplified models as the bias \cite{pawar2022multi,pawar2021physics,pawar2021model}, adopting transfer learning mechanisms \cite{de2020transfer,goswami2020transfer,chakraborty2021transfer}, designing composite networks \cite{meng2020composite}, implementing physics-informed neural networks \cite{raissi2019physics,karniadakis2021physics} and residual forcing \cite{garg2022physics}, imposing conservation laws of physical quantities or analytical constraints into neural network \cite{mohan2020embedding,beucler2021enforcing,greydanus2019hamiltonian}, and embedding tensorial invariance and equivariance properties \cite{ling2016reynolds,zanna2020data,novati2021automating,tai2019equivariant}.

Alternatively, projection-based reduced order models (PROMs) can be viewed as a physics-constrained ML methodology to emulate the system's dynamics. In particular, an effective low rank subspace is identified by means of modal analysis techniques that tailor a set of basis functions or modes representative of the dominant recurrent structures. The underlying physical constraints are imposed by performing a Galerkin projection of the governing equations onto the respective basis functions. To ensure computational efficiency, only a few basis functions are retained to build the Galerkin reduced order model (GROM). The combination of modal decomposition and projection techniques have been widely applied to build lightweight computational models in flow control systems \cite{noack2011reduced,brunton2015closed,pastoor2008feedback}. Nonetheless, the number of required modes to sufficiently describe systems of interest can be quite large. This is especially true for systems with strong nonlinearity or extreme variations in the parameter space. For such, the GROM fails to accurately represent the system's trajectory. Moreover, GROM can yield long-term instabilities even if the original system is stable \cite{ahmed2021closures}. Therefore, correcting the GROM dynamics by introducing closure terms, stabilization schemes, or regularizers is a critical step to adopt them in a reliable framework. 

The closure problem has been studied extensively in the fluid dynamics and flow control community. Structural and/or functional relationships  are often postulated, then physical and mathematical arguments are imposed to define the required parameterization. Alternatively, we address the closure modeling problem by viewing its effect as a control input applied in the latent space (i.e., latent control or latent action) to counteract the induced instabilities and inaccuracies from the GROM truncation. In particular, we employ a continuous time control signal to correct and stabilize the GROM trajectory by deriving low-rank closure models using principles from the Kolmogorov energy cascade of turbulence and energy conservation. We utilize the forecast error, measured as the discrepancy between GROM predictions and collected sensor data, as the feedback and develop a variational approach to update the control input. In addition, we leverage the forward sensitivity method (FSM) to derive first-order estimates of the relationships between the feedback and the desired control parameters \cite{lakshmivarahan2010forward}. 

\textcolor{rev}{When dealing with deterministic models, whether they are continuous or discrete, the FSM approach can be employed to effectively rectify the forecast errors that arise from inaccuracies in the initial conditions, boundary conditions, and model parameters (collectively called control) \cite{lakshmivarahan2017forecast}. Specifically, the FSM framework possesses a significant benefit, which is its independence from a backward adjoint formulation. Instead, it transforms a dynamic data assimilation problem into a static, deterministic inverse problem, thereby constituting the primary tenet of this approach. The FSM technique employs a linear Taylor series approximation to derive sensitivity dynamics for control parameters, subsequently facilitating the translation of recurrence matrix equations for forward sense. While the computation of these recurrence relations may be computationally intensive for high dimensional state problems, the FSM method holds considerable appeal for models based on latent space projection.}

\textcolor{rev}{Our approach addresses the challenge of closure problems in under-resolved regimes by using a novel strategy to optimize control input parameters at the ROM level, which allows for a more accurate description of complex physical phenomena using a small number of modes.} We highlight that one key aspect of the proposed framework is its flexibility in dealing with state variables and observables that live in distinct spaces. For example, the original system has a high dimensional state variable that lives in the physical space. In contrast, the reduced order system has a latent state variable defined in a low rank subspace. Finally, the observable output can be a different measurable quantity related to either space. This is conceptually related to the reduced order observers \cite{dada2020generalized} and functional observers \cite{kravaris2016functional,sadamoto2013low,niazi2019scale} developed in the control community. We demonstrate the proposed framework using the semi-discretized high dimensional flow problems corresponding to the Bateman–Burgers system and vortex merger at a large Reynolds number for the sensor data-rich and data-sparse regimes. 

This paper is organized as follows. In \cref{sec:rom}, we introduce the key elements of building GROM for high dimensional dynamical systems using a combination of proper orthogonal decomposition (\cref{sub:pod}) and Galerkin projection (\cref{sub:gp}). The closure problem is formally defined in \cref{sec:closure} and the proposed FSM-based control approach is presented in \cref{sec:fsm}. Numerical experiments are provided in \cref{sec:res} with the corresponding discussions. Finally, \cref{sec:conc} draws the main conclusions of the study and offers outlook for future work.

\section{Galerkin Reduced Order Models} \label{sec:rom}
We consider an autonomous dynamical system defined as follows:
\begin{equation}
    \textcolor{rev}{\dot{\bu} = \mathcal{F}(\bu),} \label{eq:fom}
\end{equation}
where $\bu \in \mathbb{R}^N$ is the state vector (e.g., the value of the velocity field at discrete grid points) and $\mathcal{F}:\mathbb{R}^N \times \to \mathbb{R}^N$ represents the system's dynamics (e.g., the spatial discretization of the Navier-Stokes equations). We note that \cref{eq:fom} is often called the full order model (FOM) or high dimensional model (HDM) in ROM studies. Due to the computational complexity of solving \cref{eq:fom} for large scale systems with millions of degrees of freedom (DOFs), FOMs are not feasible for multi-query applications (e.g., inverse problem and model predictive control). A possible mitigation strategy is to replace the state vector $\bu$ with a lower rank approximation, where the solution is approximated using a few basis functions that capture the main characteristics of the system. 
\subsection{Proper Orthogonal Decomposition} \label{sub:pod}
Proper orthogonal decomposition (POD) is one of the modal decomposition techniques that has been used successfully over last few decades to define optimal low rank bases for the quantities of interest \textcolor{rev}{\cite{sirovich1987turbulence,aubry1991hidden,berkooz1993proper,holmes2012turbulence,cordier2013identification,lassila2014model,taira2017modal,taira2020modal}}. The POD procedure begins with a set of pre-collected realizations of the system's behavior (known as flow snapshots) at different times as follows: 
\begin{equation}
\mathcal{U} := \{ \bu^{(1)}, \bu^{(2)}, \dots, \bu^{(K)} \}, \label{eq:snap}
\end{equation}
\textcolor{rev}{where $\bu^{(i)}$ denotes the $i^{\text{th}}$ snapshot reshaped into a column vector.} A Reynolds decomposition of the flow field $\bu$ can be written as:
\begin{equation}
    \bu = \bar{\bu} + \bu',
\end{equation}
where $\bar{\bu}$ is a reference field usually defined by the ensemble mean as follows:
\begin{equation}
    \bar{\bu} = \dfrac{1}{K} \sum_{i=1}^{K} \bu^{(i)},
\end{equation}
and thus $\bu'$ represents the fluctuating component of the field. POD (using the method of snapshots) seeks a low rank basis functions for the span of $\mathcal{U}' := \{ \bu'^{(1)}, \bu'^{(2)}, \dots, \bu'^{(K)} \}$ by defining a correlation matrix $\mathbf{C} \in \mathbb{R}^{K\times K}$ as follows:
\begin{equation}
    [\mathbf{C}]_{ij} = (\bu'^{(i)}, \bu'^{(j)}),
\end{equation}
where $(\cdot, \cdot)$ denotes the appropriate inner product. An eigenvalue decomposition of $\mathbf{C}$ yields a set of eigenvectors $\mathbf{V} = [\mathbf{v}_1, \mathbf{v}_2, \dots \mathbf{v}_K]$ and the corresponding eigenvalues $\boldsymbol{\Lambda} = \text{diag}[\lambda_1, \lambda_2, \dots \lambda_K]$ as: 
\begin{equation}
    \mathbf{C} \mathbf{V} = \mathbf{V} \boldsymbol{\Lambda}.
\end{equation}
For optimal basis selection, the eigenvalues are stored in descending order \textcolor{rev}{of magnitude} (i.e., $\lambda_1 \ge \lambda_2\ge \dots \ge \lambda_K \ge 0$). The POD basis functions $\boldsymbol{\Phi} = \{\bphi_1, \bphi_2, \dots \bphi_K\}$ can be recovered as follows:
\begin{equation}
    \textcolor{rev}{\bphi_i = \dfrac{1}{\sqrt{\lambda_i}} \sum_{j=1}^{K} \mathbf{v}_{i,j} \bu'^{(j)},} \label{eq:phi}
\end{equation}
\textcolor{rev}{where $\mathbf{v}_{i,j}$ is the $j^{\text{th}}$ component of the $i^{\text{th}}$ eigenvector.} Finally, the $n^{\text{th}}$ rank POD approximation of $\bu$ is obtained by considering only the first $n$ basis functions as follows:
\begin{equation}
    \bu \approx \bar{\bu} + \sum_{i=1}^{n} a_i \bphi_i. \label{eq:upod}
\end{equation}

\subsection{Galerkin Projection} \label{sub:gp}
In \cref{eq:upod}, the mean field $\bar{\bu}$ and the basis functions $\boldsymbol{\Phi}$ are computed from the collected snapshots during an offline stage. In order to estimate $\bu$ at arbitrary times and/or parameters, a model that describes the variation of the coefficients $\ba = [a_1, a_2, \dots, a_n]^T$ is required. The Galerkin ROM (GROM) of the dynamical system governed by \cref{eq:fom} is obtained by replacing $\bu$ by its $n^{\text{th}}$ rank POD approximation from \cref{eq:upod}, followed by an inner product with arbitrary POD modes to yield the following system of ordinary differential equations:
\begin{equation}
    \dot{a}_k = \bigg( \mathcal{F}(\bar{\bu} + \sum_{i=1}^{n} a_i \bphi_i), \bphi_k\bigg), \quad \text{for } k=1,2,\dots, n.  \label{eq:rom2}
\end{equation}
We note that the simplification in the left hand-side of \cref{eq:rom2} takes advantage of the orthornomality of the POD basis function (i.e., $(\bphi_i,\bphi_j) = \delta_{ij}$ where $\delta_{ij}$ is the Kronecker delta). \textcolor{rev}{Without loss of generality, we consider the following incompressible Navier-Stokes equation (NSE) for demonstration purposes:
\begin{equation}
    \begin{aligned}
        \dfrac{\partial \bu}{\partial t}  + (\bu \cdot \nabla) \bu  &=  - \nabla p + \nu \Delta \bu,  \\
        \nabla \cdot \bu &= 0,
    \end{aligned} \label{eq:nse}
\end{equation}
where $\bu$ is the velocity vector field, $p$ the pressure field, and $\nu$ is the kinematic viscosity.}
This form of the NSE captures the main characteristics of a large class of flow problems with quadratic nonlinearity and second order dissipation. In our results section, we showcase the applicability of the presented approach in the 1D Burgers problem and the 2D vortex-merger flow problem governed by the vortex transport equations. Applying the Galerkin method \textcolor{rev}{to} \cref{eq:nse}, the resulting GROM reads as follows:
\begin{equation}
    \dot{\ba} = \boldsymbol{b} +  \boldsymbol{L} \ba +  \ba^T \boldsymbol{N} \ba, \label{eq:grom}
\end{equation}
where $\boldsymbol{b} \in \mathbb{R}^n$, $\boldsymbol{L} \in \mathbb{R}^{n\times n}$, and $\boldsymbol{N} \in \mathbb{R}^{n\times n \times n}$ respectively represent the constant, linear, and nonlinear terms that result from the inner product between the FOM operators and the POD basis functions.  


\section{The Closure Problem} \label{sec:closure}
Due to the modal truncation (i.e., using $n\ll K$ in \cref{eq:upod}), the effects of the truncated scales onto the resolved scales are not captured by \cref{eq:rom2}. Therefore, the resulting GROM fails to accurately represent the dynamics of the ROM variables $\ba$. Previous studies have shown that GROM yields inaccurate and sometimes unstable behavior even if the actual system is stable \cite{grimberg2020stability}. Therefore, efforts have been focused onto developing techniques to correct the GROM trajectory, including works to correct and/or stabilize the GROM using closure and/or regularization schemes \cite{snyder2022reduced}. As introduced in \cref{sec:intro}, we view the \textcolor{rev}{process} of adjusting the GROM trajectory as a control task \textcolor{rev}{with a \emph{computational actuator} in the latent space} of the reduced order model.

To this end, we modify \cref{eq:grom} by adding a control input $\bc(t) = [c_1, c_2, \dots, c_n]^T$ as follows:
\begin{equation}
    \dot{\ba} = \boldsymbol{b} +  \boldsymbol{L} \ba +  \ba^T \boldsymbol{N} \ba + \bc(t). \label{eq:grom_c1}
\end{equation}
The goal of the control $\bc$ is to steer the GROM predictions toward the target trajectory defined as follows:
\textcolor{rev}{
\begin{equation}
    \widehat{a}_k(t) = \bigg( \bu(t) - \bar{\bu}, \bphi_k\bigg), \label{eq:atrue}
\end{equation}
where the superscript $\widehat{(\cdot)}$ denotes the target values.} It can be verified that the trajectory given in \cref{eq:atrue} with the POD basis functions $\bphi_k$ yields the minimum approximation error among all possible reconstructions of rank-$n$ (or less) \cite{holmes2012turbulence}. The control input that would result in values of $\ba$ that are exactly equal to their optimal values in \cref{eq:atrue} can be defined as follows:
\begin{equation}
     c_k(t) = \bigg(\mathcal{F}(\bu), \bphi_k\bigg) -  \bigg( \mathcal{F}(\bar{\bu} + \sum_{i=1}^{n} a_i \bphi_i), \bphi_k\bigg). \label{eq:ctrue}
\end{equation}
\Cref{eq:ctrue} essentially leads to the following:
\begin{equation}
     \dot{a}_k = \bigg(\mathcal{F}(\bu), \bphi_k\bigg), \label{eq:adottrue}
\end{equation}
which is an exact equation (i.e., no truncation) for the dynamics of $\ba$. However, we highlight that \cref{eq:ctrue} is not useful in practice as it requires solving the FOM to compute $\bu$ at each time step. Therefore, alternative approximate models are sought to estimate $\bc$ as a function of the available information in the ROM subspace (i.e., $\{a_k,\phi_k\}_{k=1}^{n}$). To account for the effect of ROM truncation onto the dynamics of ROM scales themselves is often referred to as the \emph{closure problem}.

The development of closure models for ROMs has been largely influenced by turbulence modeling and especially large eddy simulation (LES) studies. For example, by analogy between POD modes and Fourier modes, it is often postulated that the high-index modes (i.e., $\{\phi_k\}_{k>n}$) are responsible for dissipating the energy. In turn, by truncating these modes, energy accumulates in the systems causing instabilities. In this regard, the addition of artificial dissipation through eddy viscosity has shown substantial success in improving ROM accuracy \cite{borggaard2011artificial,akhtar2012new,wang2012proper,san2014proper}. Nonetheless, the determination of the eddy viscosity term has been a major challenge. Several studies relied on brute-force search to select optimal values while other works were inspired by state-of-the-art LES models \cite{ahmed2021closures}, such as the Smagorinksy model \cite{akhtar2012new,san2015stabilized} or its dynamic counterparts \cite{wang2012proper,rahman2019dynamic}. Noack et al. \cite{noack2011reduced} utilized a finite-time thermodynamics approach to quantify a nonlinear eddy viscosity by matching the modal energy transfer effect. A notably distinct closure model was proposed in \cite{cazemier1998proper} by adding a linear damping term to the ROM equation. This model is predefined using the collected ensemble of snapshots following an energy conservation analysis. 

The present study draws concepts from the Kolmogorov energy cascade and energy conservation principles to define the effect of the modal truncation on ROM dynamics. \textcolor{rev}{In order to derive the form of the closure model, we add a combination of linear friction and diffusion terms to \cref{eq:nse} as follows:
\begin{equation}
    \dfrac{\partial \bu}{\partial t}  + (\bu \cdot \nabla) \bu  = - \nabla p + \nu \Delta \bu  + \gamma \bu + \beta  \Delta \bu, \label{eq:nse_c}
\end{equation}
where $\gamma$ and $\beta$ are the friction and diffusion parameters, respectively. Projecting \cref{eq:nse_c} onto the POD subspace leads to a model for $\bc$ as follows:
\begin{align}
    \bc(t) &= \gamma \boldsymbol{e} + \gamma \ba + \beta \boldsymbol{q} + \beta \boldsymbol{D} \ba \label{eq:closure} \\
\text{where: } \qquad 
    \relax [\boldsymbol{e}]_{k} &= \big( \bar{\bu}, \bphi_k\big), \quad
    [\boldsymbol{q}]_{k} = \big( \Delta \bar{\bu}, \bphi_k\big), \quad
    [\boldsymbol{D}]_{k,i} = \big( \Delta \bphi_i, \bphi_k\big). \quad
\end{align}
Thus, we aim at correcting the GROM trajectory by estimating optimal values for $\gamma$ and $\beta$ and we refer to them as the \emph{control parameters} or simply the \emph{control}. However, we highlight that even if we hypothesize that the correction term can be \emph{approximated} using \cref{eq:closure}, this approximation is by no means exact. Instead, we are interested in values of $\alpha$ and $\beta$ that would result in the least error (with respect to a set of reference data points).} In addition, it should be noted here that inconsistency issues (between the full order model and reduced order model) might arise from the introduction of arbitrary closure models. We refer the interested readers to \cite{pacciarini2014stabilized,giere2015supg,stabile2019reduced,strazzullo2022consistency}. In the present study, we set $\gamma = [\gamma_1, \gamma_2, \dots, \gamma_n]^T \in \mathbb{R}^n$ and $\beta = [\beta_1, \beta_2, \dots, \beta_n]^T \in \mathbb{R}^n$ to allow variability of the closure model with different modes. The use of mode-dependent correction has been shown to provide better closure models, e.g., by matching energy levels between FOM and ROM \cite{rempfer1993dynamics}, incorporating spectral kernels \cite{sirisup2004spectral,san2014proper}, or utilizing the variational multiscale framework \cite{wang2012proper,iliescu2014variational,eroglu2017modular}.

\section{Forward Sensitivity Method} \label{sec:fsm}
We leverage the forward sensitivity method (FSM) \cite{lakshmivarahan2010forward,lakshmivarahan2017forecast} to estimate the parameters $\gamma$ and $\beta$ from a combination of the underling dynamical model and collected observational data. To simplify our notation, we rewrite \cref{eq:grom_c1}, with $\bc(t)$ defined using \cref{eq:closure}, as follows:
\begin{equation}
    \dot{\ba} = \bff(\ba,\btau), \label{eq:model}
\end{equation}
where 
\begin{equation}
    \bff(\ba,\btau) = \boldsymbol{b} +  \boldsymbol{L} \ba +  \ba^T \boldsymbol{N} \ba + \gamma \boldsymbol{e} + \gamma \ba + \beta \boldsymbol{q} + \beta \boldsymbol{D} \ba
\end{equation}
and 
\begin{equation}
\btau = [\gamma_1, \gamma_2, \dots, \gamma_n, \beta_1, \beta_2, \dots, \beta_n]^T \in \mathbb{R}^{2n}
\end{equation}
denotes the control parameters. In what follows, we use a set of collected, possibly sparse and noisy, measurements to approximate how the predictions of \textcolor{rev}{\cref{eq:model}} deviate from their target values. In addition, we describe how these predictions depend on the control parameter $\btau$ in \cref{sub:sens}. Finally, we fuse these two pieces of information to derive a relationship between the model predictions, temporal measurements, and the corrected parameter values in \cref{sub:param}.

\subsection{Forecast Error and Feedback} \label{sub:obs}
We monitor the model behavior by collecting a set of measurements $\bz \in \mathbb{R}^m$ as follows:
\begin{equation}
    \bz(t) = \bh(\widehat{\ba(t)}) + \eta(t), \label{eq:measurement}
\end{equation}
where $\bh(\cdot): \mathbb{R}^n \to \mathbb{R}^m$ represents the observational operator, \textcolor{rev}{$\widehat{\ba}$ is the true value of the model state $\ba$, and $\eta$ is the sensor measurement noise. We note that the measurement $\bz$ is a function of the ground truth and the observation operator can involve a sampling operation, an interpolation, or even a mapping between different spaces. Therefore, the dimensionality $m$ of the observation $\bz$ is not necessarily equal to the dimensionality $n$ of the state $\ba$.} 

For the measurement noise, we consider a white Gaussian perturbation (i.e., $\eta(t) \sim \mathcal{N}(\mathbf{0},\mathbf{R}(t))$, where $\mathbf{R}(t)$ denotes the measurement noise covariance matrix). In most cases, $\mathbf{R}(t)$ is a diagonal matrix implying that the measurement noise from different sensors are uncorrelated to each other. For simplicity, we assume that $\mathbf{R}(t)= \sigma^2 \mathbf{I}_m$, where $\sigma$ is the standard deviation for the measurement noise and $\mathbf{I}_m$ is the $m\times m$ identity matrix. Due to deviations between the target trajectory (corresponding \cref{eq:atrue}) and the GROM solution, the resulting forecast error can be written as follows:
\begin{equation}
    \bepsilon(t) = \bz(t) - \bh(\ba(t)). \label{eq:forecast}
\end{equation}

The definition of the operator $\bh(\cdot)$ is an important component of the whole setup. One option is to stick to the fact that measurements are often collected in the physical space and thus define the forecast error in this space. Considering the Burgers problem with a velocity field $\bu \in \mathbb{R}^N$, we might be able to collect only data for $\bv \in \mathbb{R}^M$ that is related to $\bu$ as $\bv = \mathcal{H}(\bu)$, \textcolor{rev}{where $\mathcal{H}(\bu)$ is an observation operator in the physical space of $\mathbf{u}$, compared to $\bh(\ba)$ that is applied in the latent space of $\ba$}. Thus, by setting $\bz = \bv$ (i.e., $m=M$), we have $\bh(\ba) = \mathcal{H}\big(\bar{\bu}+\sum_{i=1}^{n} a_i \phi_i\big)$. Another computationally attractive option is to define the forecast error in a latent space defined as follows:
\begin{equation}
    \bv = \sum_{i=1}^m z_i \bpsi_i,
\end{equation}
where $\{\bpsi_i\}_{i=1}^m$ is some low rank basis for $\bv$ \cite{pawar2022equation}. Considering a POD basis $\bpsi$, the components of $\bz$ can be computed by projecting the field $\bv$ on the respective basis functions as $z_i(t) = \big( \bv(t), \bpsi_i\big)$. Therefore, the observation operator $\bh(\cdot)$ can be defined as follows:
\begin{equation}
    [\bh(\ba)]_{k} = \bigg( \mathcal{H}\big(\bar{\bu}+\sum_{i=1}^{n} a_i \phi_i\big), \bpsi_k\bigg).
\end{equation}

\subsection{Sensitivity Dynamics} \label{sub:sens}
Since our objective is to relate the feedback $\bepsilon$ to the desired control parameters $\btau$, we first need to define how these parameters affect the model predictions. Thus, we define the sensitivity of the model forecast $\ba$ at any time $t$ with respect to the model's parameters $\btau$ using $\mathbf{V} \in \mathbb{R}^{n\times 2n}$ as follows,
\begin{equation}
    \textcolor{rev}{[\mathbf{V}(t)]_{ij} = \bigg[ \dfrac{\partial a_i(t)}{\partial \theta_j} \bigg].} \label{eq:sens}
\end{equation}
By differentiating the dynamical model (i.e., \cref{eq:model}) with respect to its parameters $\btau$ and using the chain rule, it can be verified that $\mathbf{V}(t)$ \textcolor{rev}{evolves} according to the following linear system:
\begin{equation}
    \dot{\mathbf{V}}(t) = \mathbf{D}_{\bff}(t) \mathbf{V}(t) +  \mathbf{D}_{\bff}^{\btau}(t), \label{eq:sensd}
\end{equation}
where $\mathbf{D}_{\bff}$ and $\mathbf{D}_{\bff}^{\btau}$ symbolize the Jacobian of the model $\bff$ with respect to the state $\ba$ and parameters $\btau$, respectively as follows:
\setcounter{MaxMatrixCols}{20}
\begin{equation}
\begin{aligned}   
    \mathbf{D}_{\bff} &= 
    \begin{bmatrix} 
    \dfrac{\partial f_1}{\partial a_1} && \dfrac{\partial f_1}{\partial a_2} && \dots && \dfrac{\partial f_1}{\partial a_n} \\ \\
    \dfrac{\partial f_2}{\partial a_1} && \dfrac{\partial f_2}{\partial a_2} && \dots && \dfrac{\partial f_2}{\partial a_n} \\ \\
    \vdots && \vdots && \ddots && \vdots \\ \\
    \dfrac{\partial f_n}{\partial a_1} && \dfrac{\partial f_n}{\partial a_2} && \dots && \dfrac{\partial f_n}{\partial a_n}
    \end{bmatrix} \in \mathbb{R}^{n\times n}, \\
    \mathbf{D}_{\bff}^{\btau} &= 
    \begin{bmatrix} 
    \dfrac{\partial f_1}{\partial \gamma_1} && \dfrac{\partial f_1}{\partial \gamma_2} && \dots && \dfrac{\partial f_1}{\partial \gamma_n} 
    &&
    \dfrac{\partial f_1}{\partial \beta_1} && \dfrac{\partial f_1}{\partial \beta_2} && \dots && \dfrac{\partial f_1}{\partial \beta_n} \\ \\
    \dfrac{\partial f_2}{\partial \gamma_1} && \dfrac{\partial f_2}{\partial \gamma_2} && \dots && \dfrac{\partial f_2}{\partial \gamma_n} 
    &&
    \dfrac{\partial f_2}{\partial \beta_1} && \dfrac{\partial f_2}{\partial \beta_2} && \dots && \dfrac{\partial f_2}{\partial \beta_n} \\ \\
    \vdots && \vdots && \ddots && \vdots && \vdots && \vdots && \ddots && \vdots \\ \\
    \dfrac{\partial f_n}{\partial \gamma_1} && \dfrac{\partial f_n}{\partial \gamma_2} && \dots && \dfrac{\partial f_n}{\partial \gamma_n} 
    &&
    \dfrac{\partial f_n}{\partial \beta_1} && \dfrac{\partial f_n}{\partial \beta_2} && \dots && \dfrac{\partial f_n}{\partial \beta_n}
    \end{bmatrix} \in \mathbb{R}^{n\times 2n}
\end{aligned}
\end{equation}
Since the initial conditions of the model sate (i.e., $\ba(0)$) is independent of the model's parameters, we can set \textcolor{rev}{$\mathbf{V}(0)=0$}. Therefore, \cref{eq:sensd} can be solved along with \cref{eq:model} to compute the model's predictions at any time as well as the sensitivity of such predictions to the model's parameters.

\subsection{Parameter Estimation} \label{sub:param}
The deviations in the GROM trajectory and closure model parameterizations are denoted $\delta \ba = \widehat{\ba} - \ba$ and $\delta \btau = \widehat{\btau} - \btau$, respectively, where the superscript $\widehat{(\cdot)}$ \textcolor{rev}{corresponds to} their target values. Thus, \cref{eq:sens} can be used to relate $\delta \ba$ to $\delta \btau$ as:
\begin{equation}
    \delta \ba(t) = \mathbf{V}(t) \delta \btau. \label{eq:da}
\end{equation}

The first order Taylor expansion of $\bh(\cdot)$ around $\ba$ can be written as  $\bh(\ba_T) = \bh(\ba) + \mathbf{D}_{\bh}(\ba) \delta \ba$, where $\mathbf{D}_{\bh}$ is the Jacobian of the observation operator $\bh$. Therefore, the forecast error $\bepsilon(t)$ in \cref{eq:forecast} can be rewritten as follows:
\begin{equation}
    \begin{aligned}
        \bepsilon(t) &= \bz(t) - \bh(\ba(t)) \\
                     &= \bh(\widehat{\ba}(t)) + \eta(t) - \bh(\ba(t)) \\
                     &= \cancel{\bh(\ba(t))} + \textcolor{rev}{\mathbf{D}_{\bh}(\ba(t)) \delta \ba(t)} + \eta(t) - \cancel{\bh(\ba(t))} \\
                     &=  \textcolor{rev}{\mathbf{D}_{\bh}(\ba(t)) \delta \ba(t) + \eta(t)}.
    \end{aligned}
\end{equation}
Thus, the deterministic component of the forecast error $\bepsilon$ is linked to the correction $\delta \btau$ through following relation:
\begin{equation}
     \bepsilon(t) = \mathbf{D}_{\bh}(\ba) \mathbf{V}(t) \delta \btau. \label{eq:eps}
\end{equation}
\Cref{eq:eps} is a linear system, in the standard form of $\mathbf{A} \mathbf{x} = \mathbf{b}$. It can be written for all time instances at which observational data are available. For example, assuming measurements are collected at $t_{1}, t_{2}, \dots, t_{T}$ we can build the following linear system:
\begin{equation}
    \underbrace{\begin{bmatrix}
    \bz(t_1) - \bh(\ba(t_1)) \\ \bz(t_2)- \bh(\ba(t_2)) \\ \vdots \\ \textcolor{rev}{\bz(t_T)} - \bh(\ba(t_T))
    \end{bmatrix}}_{\boldsymbol{\xi} \in \mathbb{R}^{Tm\times1}} =
    \underbrace{\begin{bmatrix}
    \mathbf{D}_{\bh}(\ba) \mathbf{V}(t_1) \\  \mathbf{D}_{\bh}(\ba) \mathbf{V}(t_2) \\ \vdots \\
     \mathbf{D}_{\bh}(\ba) \mathbf{V}(t_T)
    \end{bmatrix}}_{\mathbf{H}\in \mathbb{R}^{Tm\times2n}}
    \begin{bmatrix}\delta \btau \end{bmatrix}
\end{equation}
and linear system solvers can be utilized to compute optimal values for the parameters $\btau$. Furthermore, to account for the fact that the measurements, and hence the forecast errors, are subject to uncertainty, we use a weighted least-squares approach with $\mathbf{R}^{-1}$ being the weighting matrix, where $\mathbf{R}$ is a block-diagonal matrix constructed as follows,
\begin{equation}
    \mathbf{R} = \begin{bmatrix}
                 \mathbf{R}(t_1) &              &        & \\
                                & \mathbf{R}(t_2) &        & \\
                                &              & \ddots & \\
                                &              &        & \textcolor{rev}{\mathbf{R}(t_T)}
                \end{bmatrix} \in \mathbb{R}^{Tm\times Tm}.
\end{equation}
In particular, the solution of the resulting linear system can be written as follows:
\begin{equation}\label{eq:LSsolve}
\delta \btau \in \mathbb{R}^{2n\times1} = \begin{cases}
                     \left( \mathbf{H}^T \mathbf{R}^{-1} \mathbf{H} \right)^{-1} \mathbf{H}^T \mathbf{R}^{-1} \boldsymbol{\xi}, &\quad\text{over-determined: } Tm > 2n, \\
                    \mathbf{R}^{-1} \mathbf{H}^T  \left( \mathbf{H} \mathbf{R}^{-1} \mathbf{H}^T \right)^{-1} \boldsymbol{\xi}, &\quad\text{under-determined: } Tm < 2n.
                    \end{cases}
\end{equation}
\textcolor{rev}{The solution of \cref{eq:LSsolve} is repeated until convergence is obtained (i.e., no more updates to $\btau$ is needed, as shown in the psoudecode given by Algorithm 1).} 
\begin{algorithm}
  \caption{Pseudocode of the forward sensitivity method for parameter estimation.}
  \begin{algorithmic}[1]
    \State Start with the initial value of control $\btau$ and compute the model trajectory $\ba (t)$
    \State Compute sensitivity dynamics $\mathbf{V}(t)$
    \State Assemble $\mathbf{H}$ 
    \State Solve $ \mathbf{H} \delta \btau = \boldsymbol{\xi}$ as a weighted linear least squares using the weight $\mathbf{R}^{-1}$
    \State Set $\btau \leftarrow \btau + \delta \btau $
  \end{algorithmic}
\end{algorithm}

\begin{figure}[ht!]
    \centering
    \includegraphics[width=1\linewidth]{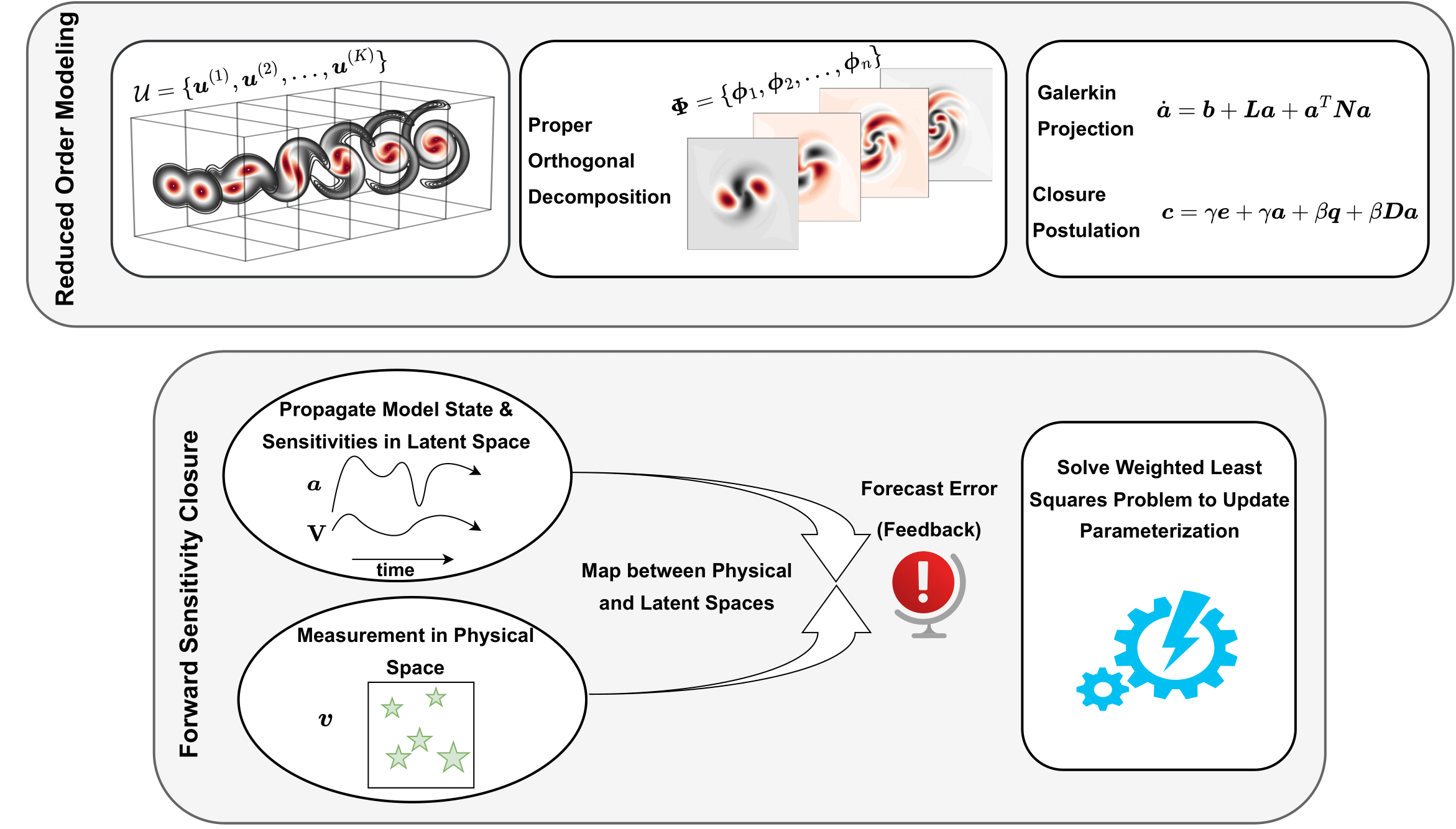}
    \caption{A schematic illustration of the algorithmic steps for deriving a Galerkin reduced order model (top) and closure modeling in the form of a latent control input using a forward sensitivity analysis (bottom).}
    \label{fig:description}
\end{figure}

\Cref{fig:description} depicts the process of identifying a lower order model for a high dimensional system and the use of the forward sensitivity framework to parameterize closure models that corrects the truncated GROM predictions. It is worth noting that once the measurement data are incorporated to estimate the parameter $\btau$, the model \cref{eq:model} can be used to make predictions at any time (i.e., not restricted to the instants when measurement data are available).

\section{Results \& Discussion} \label{sec:res}
We demonstrate the FSM closure framework using two canonical test problems with different levels of complexity. The first case deals with a one dimensional nonlinear advection diffusion system governed by the viscous Burgers equation, \textcolor{rev}{which is considered the 1D version of the NSE (see \cref{eq:nse}).} In the second demonstration, we consider the two dimensional vortex merger problem governed by the vorticity transport equation (i.e., the curl of the two dimensional NSE). We study the cases of full field measurement and the more practical scenario when only very sparse sensor data are available.

\subsection{Viscous Burgers Problem}
\begin{figure}[ht!]
    \centering
    \includegraphics[width=0.85\linewidth]{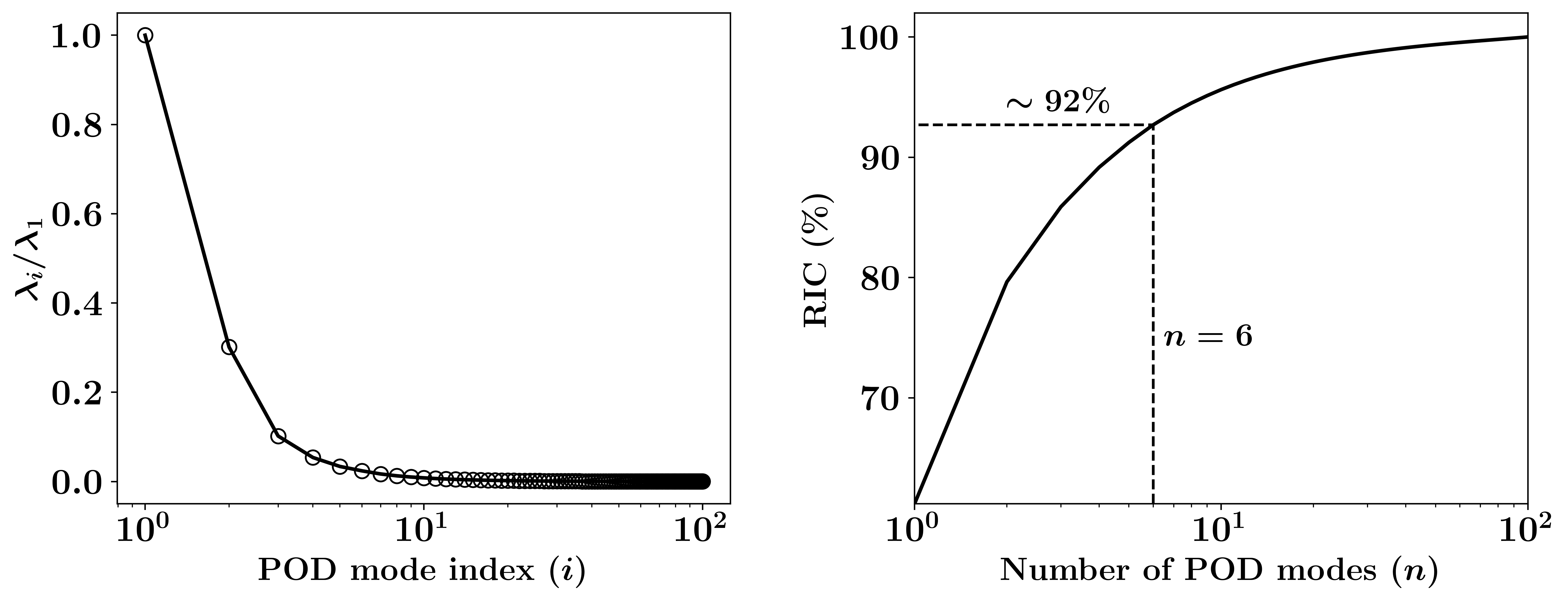}
    \caption{\textcolor{rev}{The decay of eigenvalues (left) and behavior of relative information content (right) for the 1D Burgers problem at $\text{Re}=10^4$.}}
    \label{fig:uric}
\end{figure}

\textcolor{rev}{The 1D viscous Burgers problem can be written as:
\begin{equation}
    \dfrac{\partial u}{\partial t} + u\dfrac{\partial u}{\partial x} = \nu \dfrac{\partial^2 u}{\partial x^2}. \label{eq:brg}
\end{equation}
We perform our numerical experiments at a Reynolds number $\text{Re} = 10,000$, which is equivalent to setting $\nu = 10^{-4}$ in \cref{eq:brg}.} For FOM solution, we utilize a family of compact finite difference schemes for spatial discretization and the third order total variation diminishing Runge-Kutta (TVD-RK3) scheme for temporal integration \cite{san2014proper}. We assume an initial condition of a unit step function as follows:
\begin{equation}
    \bu(x,0) = \begin{cases} 
    1, \quad \text{if} \quad x \in [0,0.5],\\
    0, \quad \text{if} \quad x \in (0.5,1].
    \end{cases}
\end{equation}
We divide the spatial domain into $4096$ equally spaced intervals and utilize a time step of $\Delta t_{FOM} = 10^{-4}$ for the FOM solution. We store velocity field snapshots every $100$ time steps to perform the POD analysis. For GROM, we retain $n=6$ modes to approximate the velocity field. An eigenvalue analysis reveals that $6$ modes capture about $92\%$ of the total system turbulent kinetic energy defined as $\frac{1}{2} \langle u_i u_i \rangle$, where $\langle \cdot \rangle$ denotes an averaging operation. \textcolor{rev}{In particular, we use the relative information content ($\text{RIC}$) metric, as shown in \cref{fig:uric} and defined as follows:
\begin{equation}
    \text{RIC}(n) = \dfrac{\sum_{i=1}^{n} \lambda_i}{\sum_{i=1}^{m} \lambda_i}\times 100.
\end{equation}
The inner product between \cref{eq:brg} and the POD basis functions leads to the GROM in \cref{eq:grom}, where the corresponding coefficients can be defined as follows:
\begin{equation}
\begin{aligned}
    [\boldsymbol{b}]_{k} &= \bigg(  \nu \dfrac{\partial^2 \bar{u}}{\partial x^2} - \bar{u}\dfrac{\partial\bar{u}}{\partial x}, \bphi_k\bigg), \\
    [\boldsymbol{L}]_{k,i} &= \bigg(  \nu \dfrac{\partial^2 \bphi_i}{\partial x^2} - \bar{u} \dfrac{\partial \bphi_i}{\partial x} - \bphi_i \dfrac{\partial \bar{u}}{\partial x}  , \bphi_k\bigg), \\
    [\boldsymbol{N}]_{k,i,j} &= \bigg(  - \bphi_i \dfrac{\partial \bphi_j}{\partial x}, \bphi_k\bigg). 
\end{aligned}
\end{equation}
The time integration of the GROM model is carried out using a time step of $\Delta t_{ROM} = 0.01$, which is $100$ times larger than $\Delta t_{FOM}$. Although $6$ POD modes represent more than $92\%$ of the total system energy, the GROM fails to correctly capture their temporal dynamics as we shall see in the following discussions. This exemplifies the need for a mechanism to correct the GROM predictions. The modified Burgers equation to admit the closure model can be written as follows:
\begin{equation}
    \dfrac{\partial u}{\partial t} + u\dfrac{\partial u}{\partial x} = \nu \dfrac{\partial^2 u}{\partial x^2} + \gamma u + \beta \dfrac{\partial^2 u}{\partial x^2}, \label{eq:brg_closure}
\end{equation}
and the resulting closure model terms in \cref{eq:closure} are
\begin{equation}
[\boldsymbol{e}]_{k} = \big( \bar{u}, \bphi_k\big), \qquad
    [\boldsymbol{q}]_{k} = \bigg( \dfrac{\partial^2 \bar{u}}{\partial x^2}, \bphi_k\bigg), \qquad
    [\boldsymbol{D}]_{k,i} = \bigg( \dfrac{\partial^2  \bphi_i}{\partial x^2}, \bphi_k\bigg). 
\end{equation}
}
In \cref{sub:full}, we use full field measurement data to parameterize such model and in \cref{sub:sparse} we deal with the sparse data regime.

\subsubsection{Full field observations} \label{sub:full}
We assume that the sensor signal is contaminated by a white Gaussian noise with zero mean and standard deviation of $0.1$, which represents $10\%$ of the peak velocity. In other words, we define $\bv = \bu + \eta$ where $\eta \sim \mathcal{N}(0,0.01\mathbf{I})$ and thus $[\bz(t)]_i = (\bv(t) - \bar{\bu}, \bphi_i )$ (see \cref{sub:obs}). We collect measurement after every 10 time integrations of the GROM (i.e., $\Delta t_{Obs} = 0.1$). 

We refer to the solution with the target trajectory (given by \cref{eq:atrue}) as prediction with ``True Closure'' notion. On the other hand, the solution of the \emph{uncontrolled} GROM (i.e., \cref{eq:grom}) is denoted as the ``No Closure'' solution. Finally, the solution of the \emph{controlled} GROM (i.e., \cref{eq:grom_c1}) with FSM used to parameterize the presumed closure model in \cref{eq:closure} is labeled as ``FSM Closure''.

\begin{figure}[ht!]
    \centering
    \includegraphics[width=0.8\linewidth]{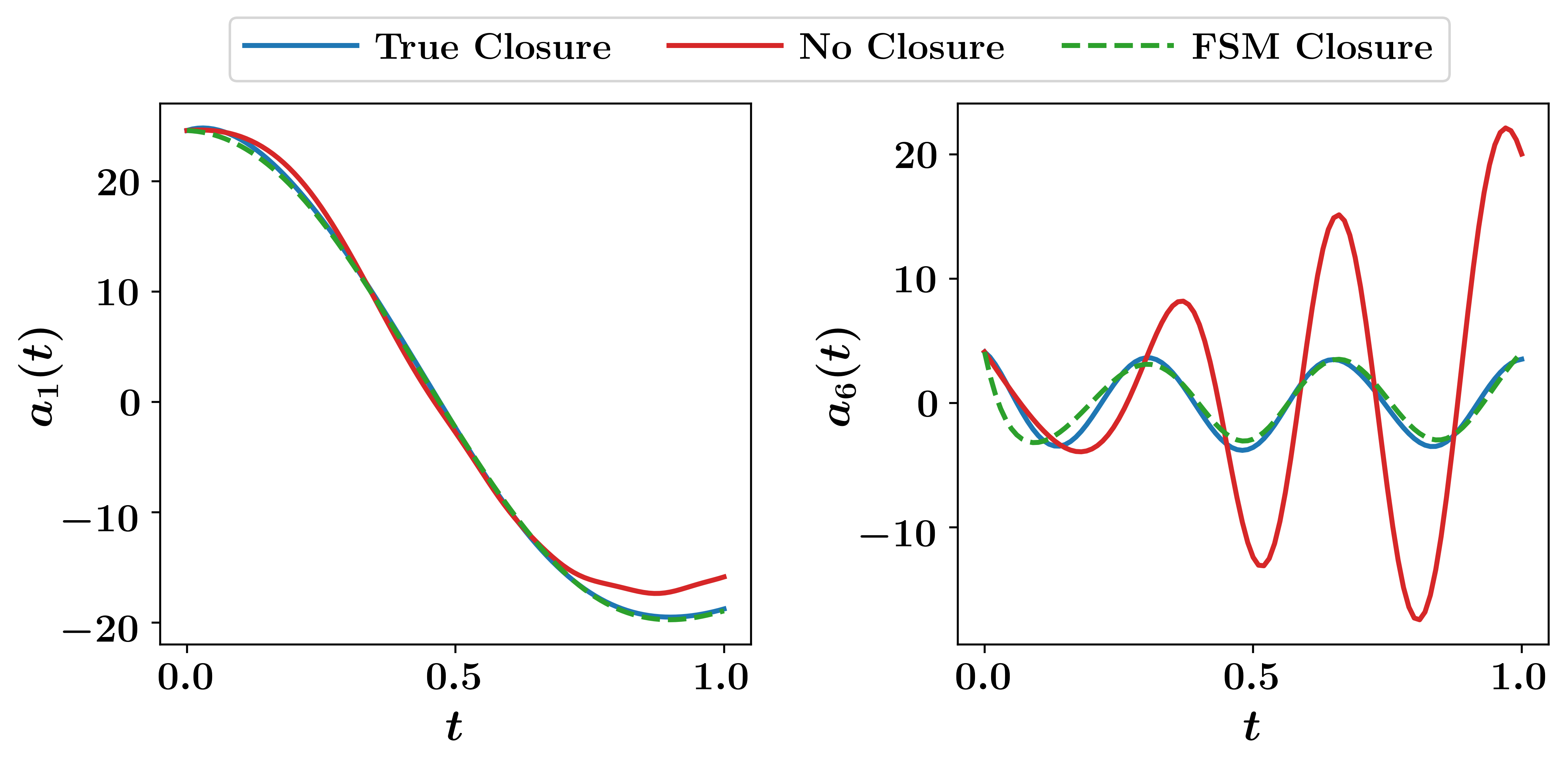}
    \caption{Dynamics of the first and last modal coefficients with full field measurement for the FSM Closure.}
    \label{fig:coeff_f}
\end{figure}

\Cref{fig:coeff_f} depicts the predicted dynamics in the latent ROM space using the considered different approaches. We observe that GROM leads to inaccuracies and significantly amplifies the magnitude of predicted coefficients, especially for the last mode. This behavior is likely to cause long term instabilities in the solution even if the actual system is stable. On the other hand, the FSM effectively controls the GROM trajectory and keeps it closer to the target trajectory. We emphasize that we implement a mode-dependent control to respect the distinct characteristics of the resolved modes defining recurrent flow structures. 

\begin{figure}[ht!]
    \centering
    \includegraphics[width=0.75\linewidth]{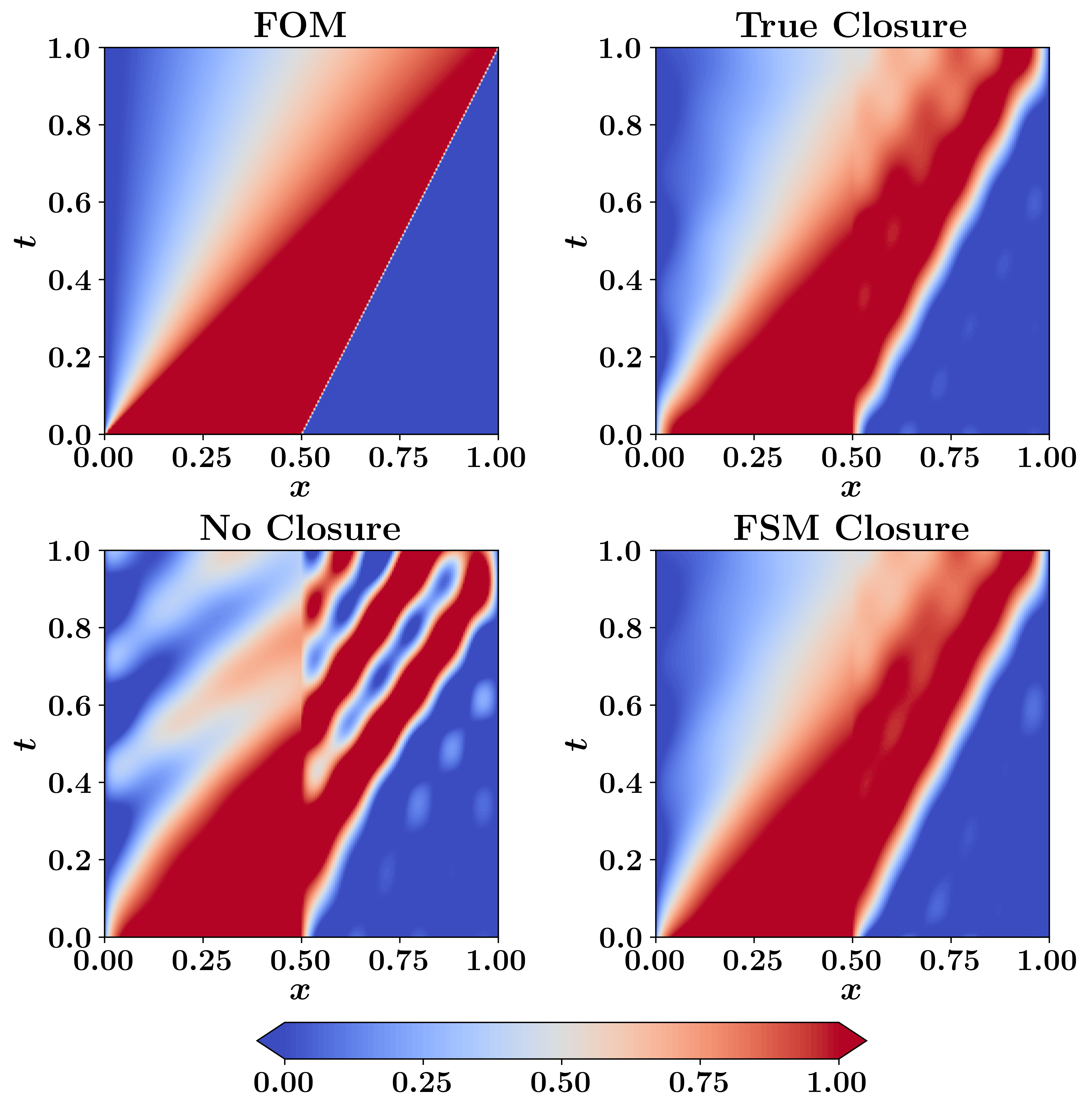}
    \caption{Spatio-temporal field predictions of Burgers problem using FOM and GROM approaches. Full field measurements are considered for the FSM Closure.}
    \label{fig:field_fc}
\end{figure}

\begin{figure}[ht!]
    \centering
    \includegraphics[width=0.85\linewidth]{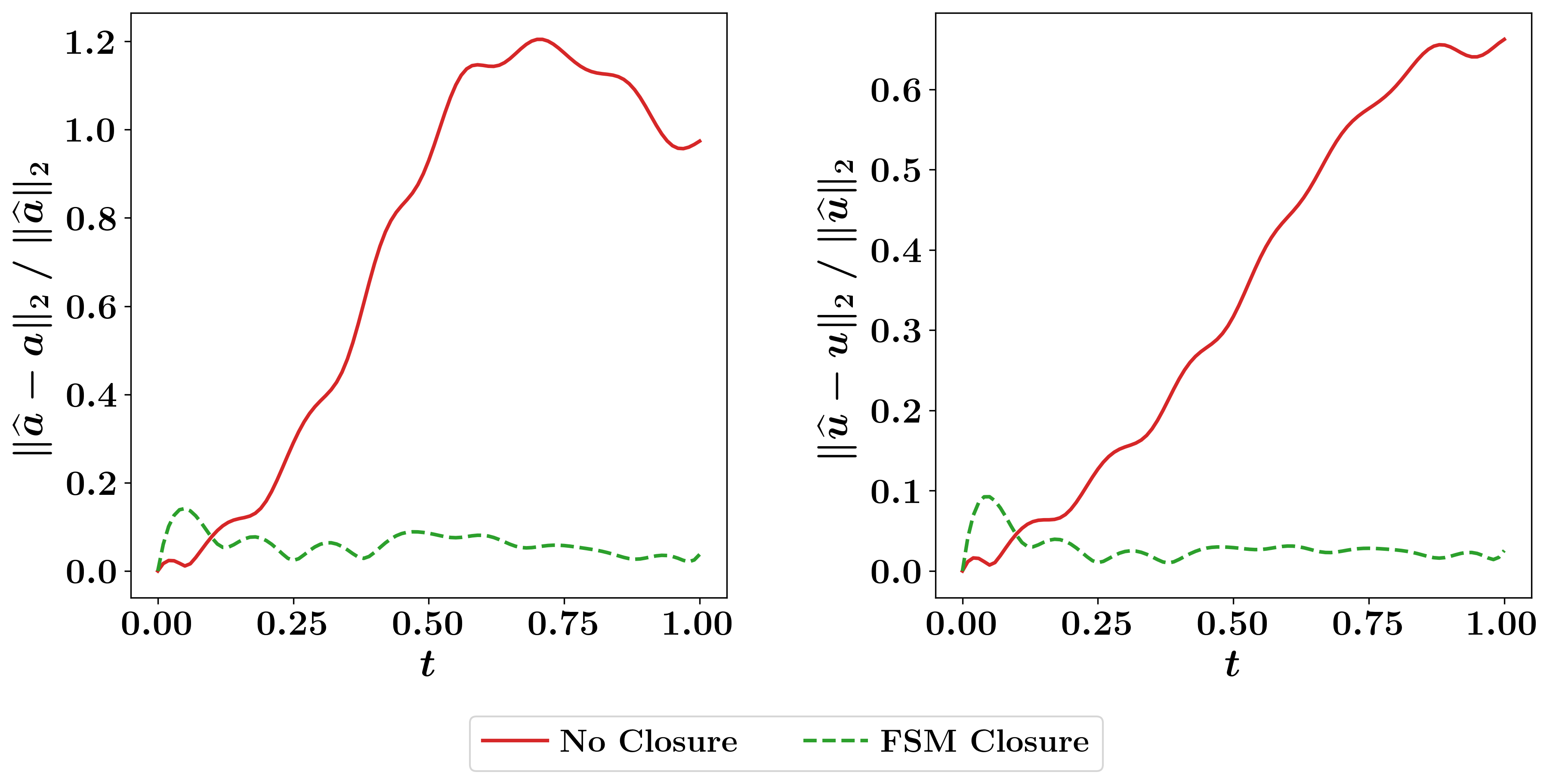}
    \caption{\textcolor{rev}{The relative error between the predicted values for the POD modal coefficients (left) and reconstructed velocity field (right) compared to their target values for 1D Burgers problem. Full field measurements are considered for the FSM Closure.}}
    \label{fig:error_fc}
\end{figure}

In \cref{fig:field_fc}, we evaluate the performance in the physical space by computing the reconstructed flow field using \cref{eq:upod} compared to the FOM fields. \textcolor{rev}{In addition, the relative error for the predicted POD coefficients as well as the reconstructed velocity fields as a function of time is shown in \cref{fig:error_fc}.} We see that results from FSM Closure are close to the True Closure which represents the minimum reconstruction error that could be obtained using $6$ modes. On the other hand, vanilla-type GROM without closure yields inaccurate and even non-physical solution in the spatio-temporal space.

\subsubsection{Sparse field observations} \label{sub:sparse}
We extend our numerical experiments to explore incomplete field measurement scenarios. In particular, we consider a sparse signal $\bs \in \mathbb{R}^S$ of the observable field $\bv$ as follows: 
\begin{equation}
    \bs = \bTheta \bv,
\end{equation}
where $\bTheta \in \mathbb{R}^{S \times M}$ is a sampling matrix, constructed by taking $S$ rows of the $M\times M$ identity matrix (i.e., $[\bTheta]_{ij} = 1$ if the $i^{th}$ sensor is located at the $j^{th}$ location and $[\bTheta]_{ij} = 0$, otherwise). Sensors can be placed at equally-spaced locations, random locations, or carefully selected places. 

Optimal sensor placement is an active field of research, also known as optimal experimental design (OED). \textcolor{rev}{We refer to \cite{alexanderian2021optimal} and references therein for more information.} In this regard, we utilize a greedy compressed sensing algorithm based on QR decomposition with column pivoting to set-up a near-optimal sensor placement strategy as follows:
\begin{equation}
    \bPsi^T \mathbf{P} := \mathbf{Q} \mathbf{R},
\end{equation}
where $\bPsi = [\bpsi_1, \bpsi_2, \dots, \bpsi_S] \in \mathbb{R}^{M\times S}$ includes the first $S$ POD basis functions for $\bv$, and $\mathbf{P} \in \mathbb{R}^{M\times M} $ is the permutation matrix. Manohar et al.~\cite{manohar2018data} showed that by using the first $S$ rows of $\mathbf{P}$ to define the sampling matrix $\bTheta$, a near optimal sensor placement is obtained with similarities to the A- and D-optimality criteria in OED studies. Finally, the field $\bv$ can be reconstructed as $\bv \approx \bPsi (\bTheta \bPsi)^{-1} \bs$. Again, if we assume that the observable $\bv$ is the velocity field $\bu$ itself, the latent measurement $\bz$ can be computed as $[\bz(t)]_i = (\bPsi (\bTheta \bPsi)^{-1} \bs(t) - \bar{\bu}, \bphi_i )$.

\begin{figure}[ht!]
    \centering
    \includegraphics[width=0.8\linewidth]{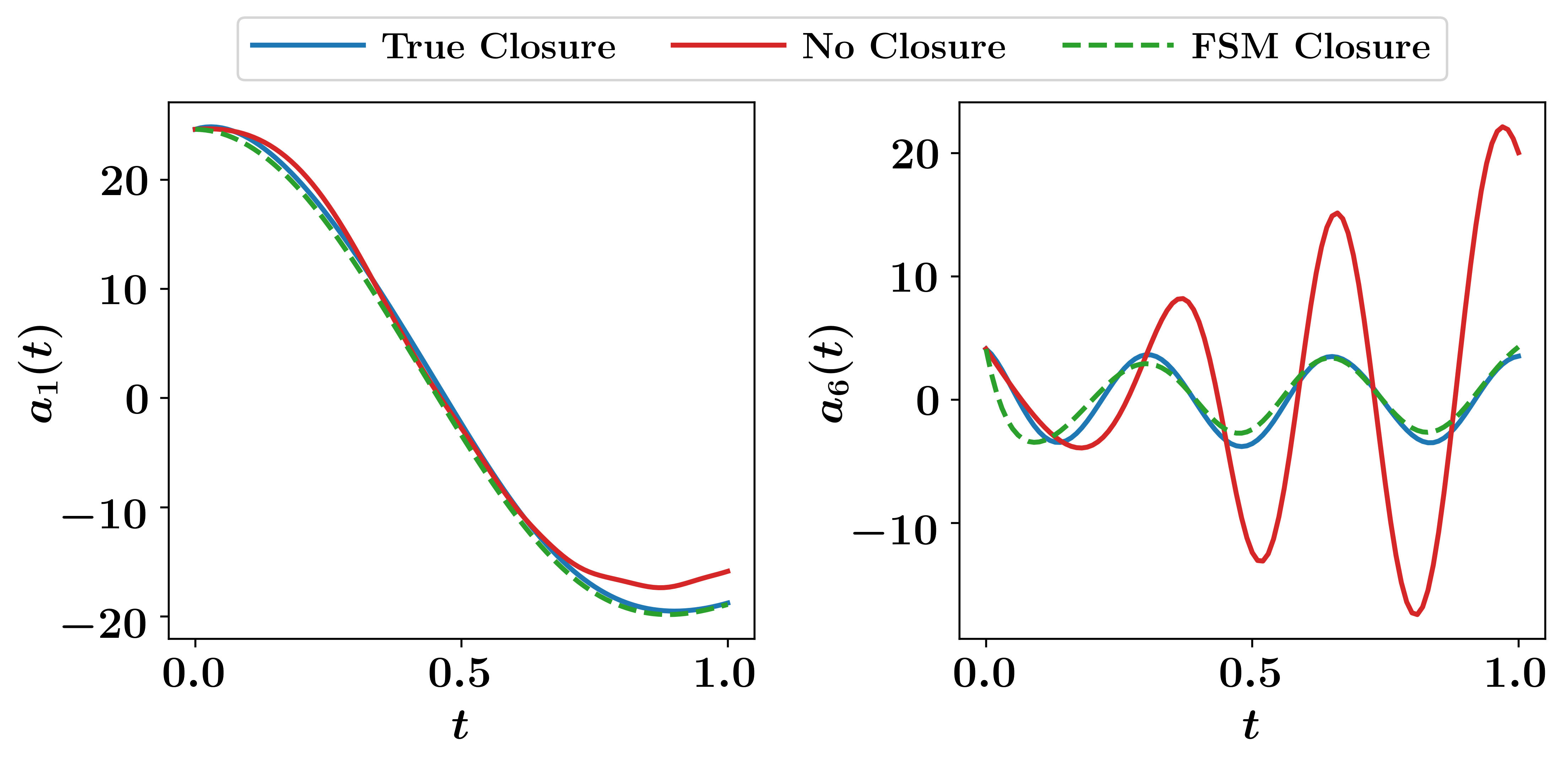}
    \caption{Dynamics of the first and last modal coefficients with sparse field measurement for the FSM Closure.}
    \label{fig:coeff_s}
\end{figure}

\begin{figure}[ht!]
    \centering
    \includegraphics[width=0.75\linewidth]{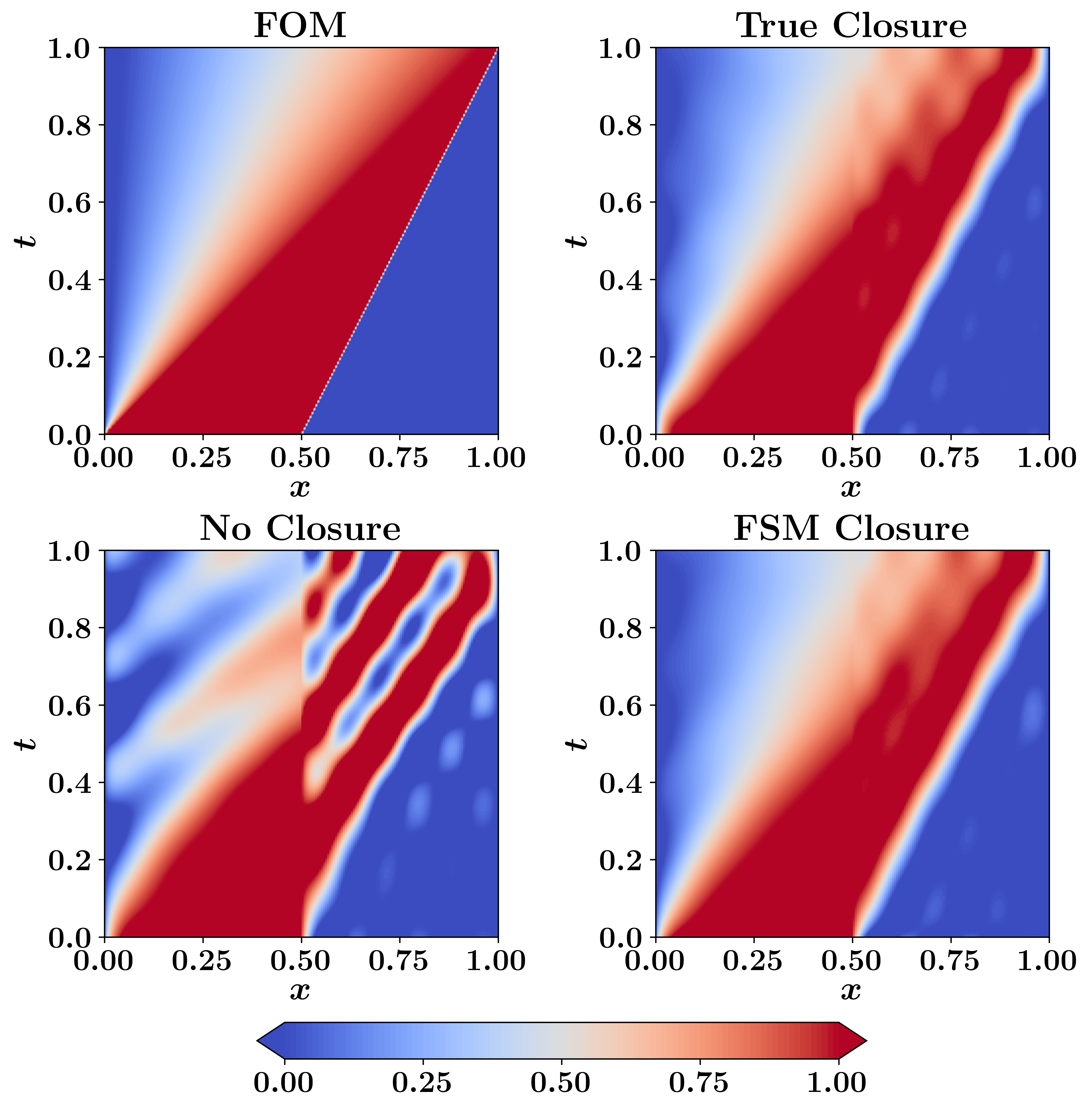}
    \caption{Spatio-temporal field predictions of Burgers problem using FOM and GROM approaches. Sparse field measurements are considered for the FSM Closure.}
    \label{fig:field_sc}
\end{figure}

\begin{figure}[ht!]
    \centering
    \includegraphics[width=0.85\linewidth]{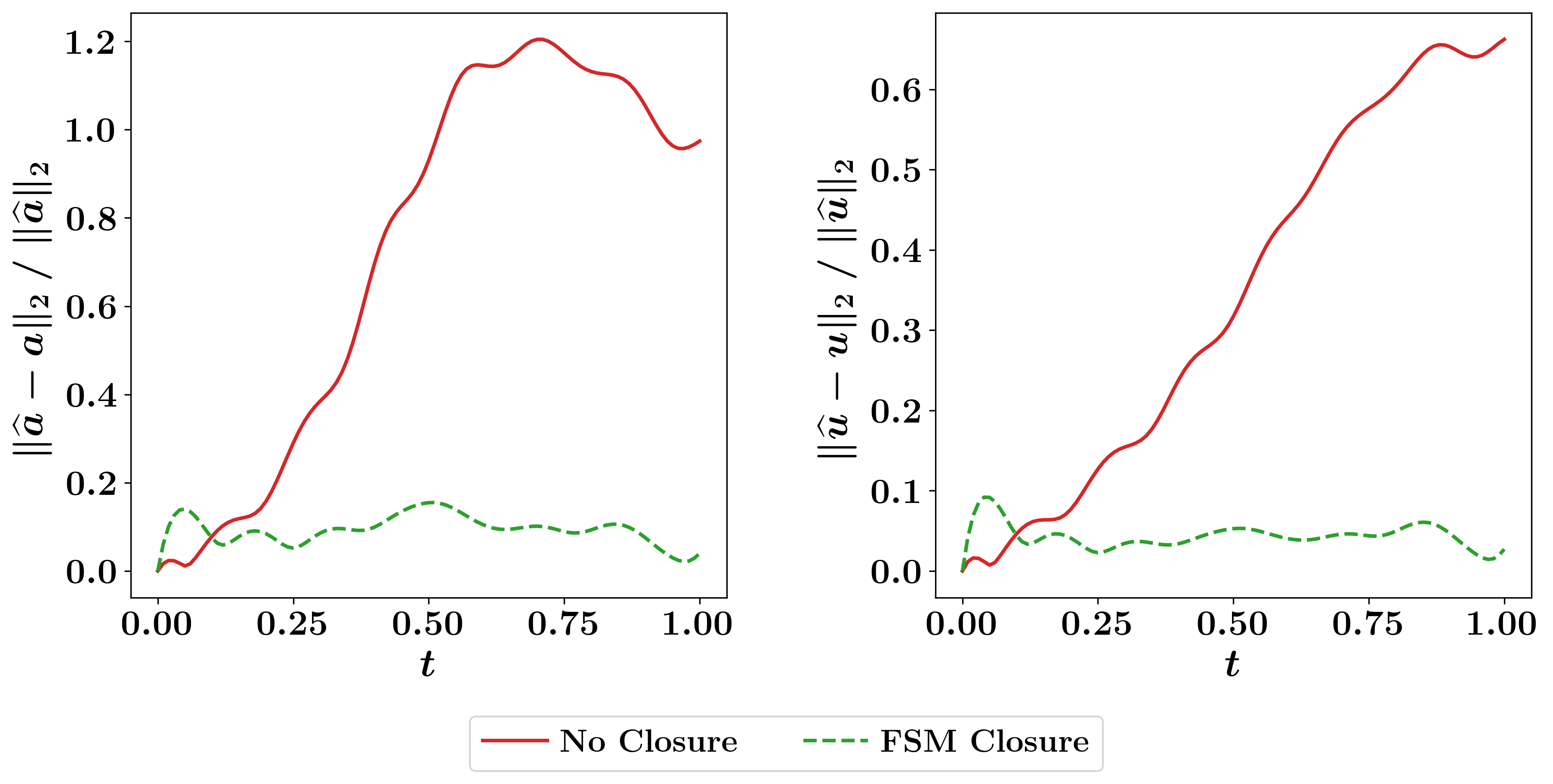}
    \caption{\textcolor{rev}{The relative error between the predicted values for the POD modal coefficients (left) and reconstructed velocity field (right) compared to their target values for 1D Burgers problem. Sparse field measurements are considered for the FSM Closure.}}
    \label{fig:error_sc}
\end{figure}

\Cref{fig:coeff_s} displays the time evolution of the first and sixth modal coefficients with the adopted FSM closure methodology in the case of sparse measurements. In particular, we selected $25$ locations (about $0.5\%$ of the total number of grid points) using the described QR-based algorithm to define the sensors data.  We see that FSM closure yields very accurate results that are close to the the target trajectory even with the sparse measurement data. The reconstruction accuracy is also demonstrated using \cref{fig:field_sc}, showing significant improvements compared the GROM predictions without control. \textcolor{rev}{Similar observations can be found in \cref{fig:error_sc} displaying the relative error for the predicted POD coefficients and the reconstructed velocity fields with respect to the target values that represent the minimum reconstruction error with $6$ modes.}

\clearpage
\subsection{Vortex Merger Problem}
One of the key benefits of the proposed FSM closure framework is that it is dealing with the reduced order model of the problem instead of the full fledged high dimensional model. Therefore, the computational complexity is dependent on the number of employed POD modes, rather than the spatial dimensionality of the problem. In order to \textcolor{rev}{highlight} this aspect, we consider the two dimensional (2D) vortex merger problem \cite{san2013coarse}, governed by the following vorticity transport equation:
\begin{equation}
\dfrac{\partial \omega}{\partial t} + J(\omega,\psi) = \dfrac{1}{\text{Re}} \Delta \omega, \qquad \text{in} \ \Omega \times [0,T]. \label{eq:vortex}
\end{equation}
where $\omega$ and $\psi$ denote the vorticity and \textcolor{rev}{streamfunction} fields, and ($J(\cdot,\cdot)$) is the Jacobian operator defined as:
\begin{align}
    J(\omega,\psi) &= \dfrac{\partial \omega}{\partial x} \dfrac{\partial \psi}{\partial y} -  \dfrac{\partial \omega}{\partial y} \dfrac{\partial \psi}{\partial x}.
\end{align}
The vorticity and \textcolor{rev}{streamfunction} are linked by the kinematic relationship:
\begin{equation}
\Delta  \psi = -\omega. \label{eq:poisson}
\end{equation}

\begin{figure}[ht!]
    \centering
    \includegraphics[width=0.9\linewidth]{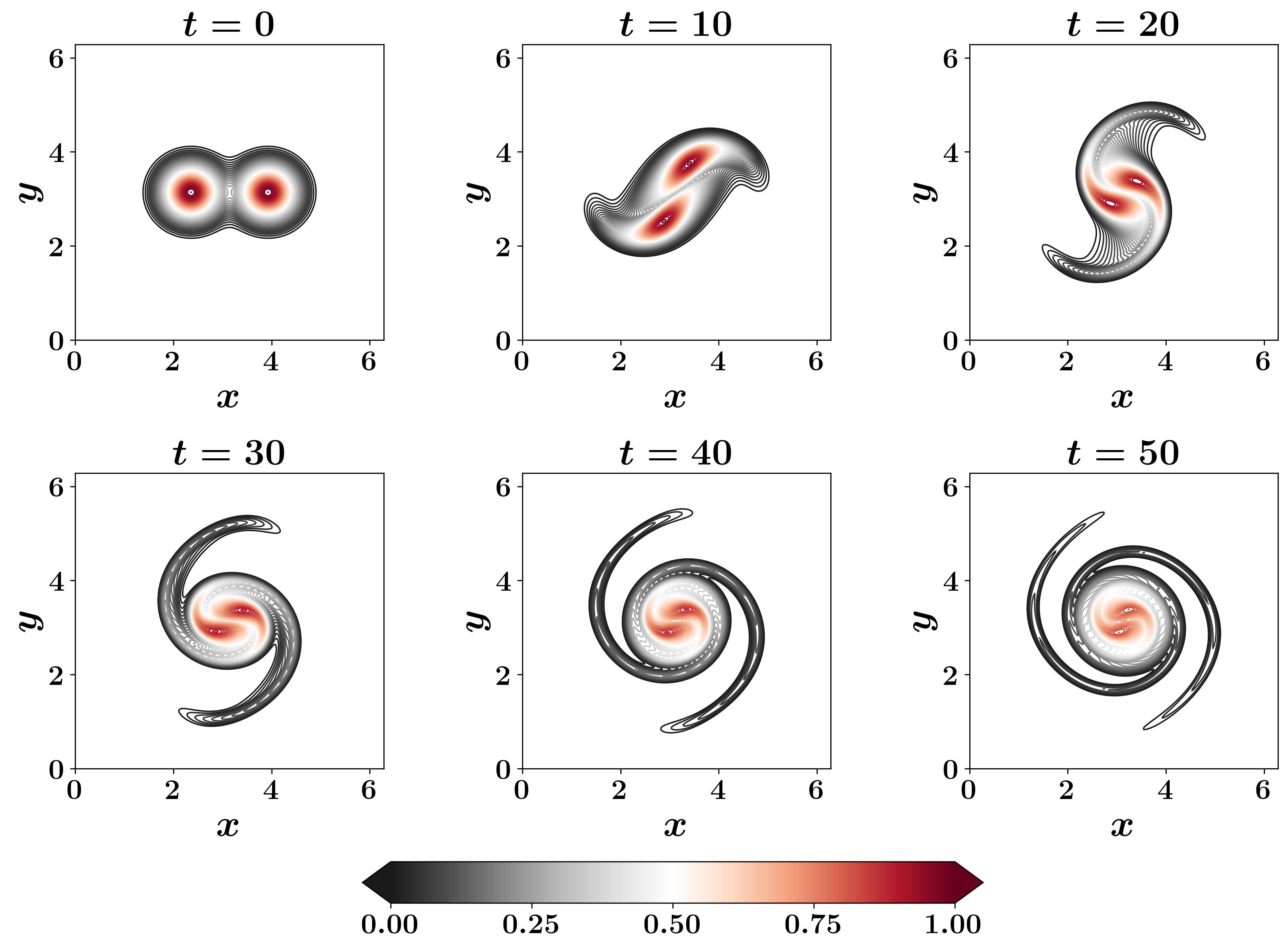}
    \caption{Samples of temporal snapshots of the vorticity field for the vortex merger problem at $\text{Re} = 5000$.}
    \label{fig:vmfom}
\end{figure}

We consider a spatial domain of dimensions $(2\pi \times 2\pi)$ with periodic boundary conditions in both the $x$ and $y$ directions. The flow is initiated with a pair of co-rotating Gaussian vortices with equal strengths centered at $(x_1,y_1) = ( 5\pi/4,\pi)$ and $(x_2,y_2) = ( 3\pi/4,\pi)$ as follows:
\begin{equation}
    \omega(x,y,0) =  \exp\left( -\rho \left[ (x-x_1)^2  + (y-y_1)^2 \right] \right) + \exp{\left( -\rho \left[ (x-x_2)^2 + (y-y_2)^2 \right] \right)}, \label{eq:vminit}
\end{equation}
where $\rho$ is a parameter that controls the mutual interactions between the two vortical motions. In the present study, we consider $\text{Re} = 5000$ and set $\rho = \pi$. For the FOM simulations, we define a regular Cartesian grid with a resolution of $256\times256$ (i.e., $\Delta x = \Delta y = 2\pi/256$). For temporal integration of the FOM model, we use the TVD-RK3 scheme with a time-step of $10^{-3}$. Vorticity snapshots are collected every 100 time-steps for $t\in [0,50]$, resulting in a total of $500$ snapshots. The evolution of the vortex merger problem is depicted in~\cref{fig:vmfom}, which illustrates the convective and interactive mechanisms affecting the transport and development of the two vortices. This makes it a challenging problem for standard ROM approaches and a good test bed for the proposed FSM framework.

\begin{figure}[ht!]
    \centering
    \includegraphics[width=0.85\linewidth]{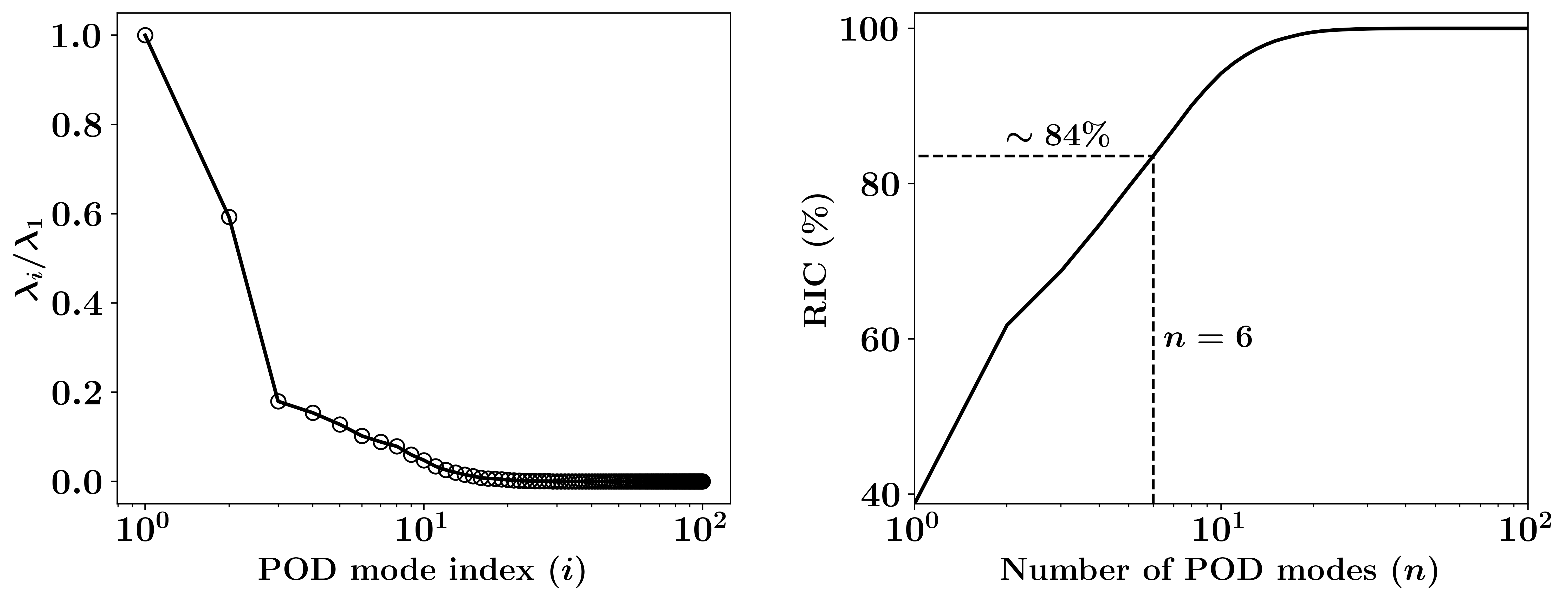}
    \caption{\textcolor{rev}{The decay of eigenvalues (left) and behavior of relative information content (right) for the 2D vortex merger problem at $\text{Re}=5000$.}}
    \label{fig:wric}
\end{figure}

In terms of POD analysis, we use $n=6$ to define the total number of resolved scales and hence the dimensionality of the GROM system. \textcolor{rev}{The decay of the POD eigenvalue and the RIC values for the current setup is shown in \cref{fig:wric}. Finally, the GROM terms for the vortex merger problem can be written as follows:
\begin{equation}
\begin{aligned}
    [\boldsymbol{b}]_{k} &= \bigg( -J(\bar{\omega},\bar{\psi}) + \dfrac{1}{\text{Re}} \nabla^2 \bar{\omega} , \bphi_k^{\omega} \bigg), \\
    [\boldsymbol{L}]_{k,i} &= \bigg(-J(\bar{\omega},\bphi_i^{\psi})  -J(\bphi_i^{\omega},\bar{\psi}) + \dfrac{1}{\text{Re}} \Delta \phi_i^{\omega} , \bphi_k^{\omega} \bigg), \\
    [\boldsymbol{N}]_{k,i,j} &= \bigg( -J(\bphi_i^{\omega},\bphi_j^{\psi}) ; \bphi_k^{\omega} \bigg).
\end{aligned}
\end{equation}
Similar to \cref{eq:brg_closure}, we modify \cref{eq:vortex} to derive the closure model as follows:
\begin{equation}
\dfrac{\partial \omega}{\partial t} + J(\omega,\psi) = \dfrac{1}{\text{Re}} \Delta \omega + \gamma \omega + \beta  \Delta \omega, \label{eq:vortex_closure}
\end{equation}
which results in the following terms for the closure model in \cref{eq:closure}:
\begin{equation}
    [\boldsymbol{e}]_{k} = \big( \bar{\omega}, \bphi_k\big), \qquad
    [\boldsymbol{q}]_{k} = \big( \Delta \bar{\omega}, \bphi_k\big), \qquad
    [\boldsymbol{D}]_{k,i} = \big( \Delta  \bphi_i, \bphi_k\big). 
\end{equation}
}

\subsubsection{Full field observations} \label{sub:vmfull}
\begin{figure}[ht!]
    \centering
    \includegraphics[width=0.9\linewidth]{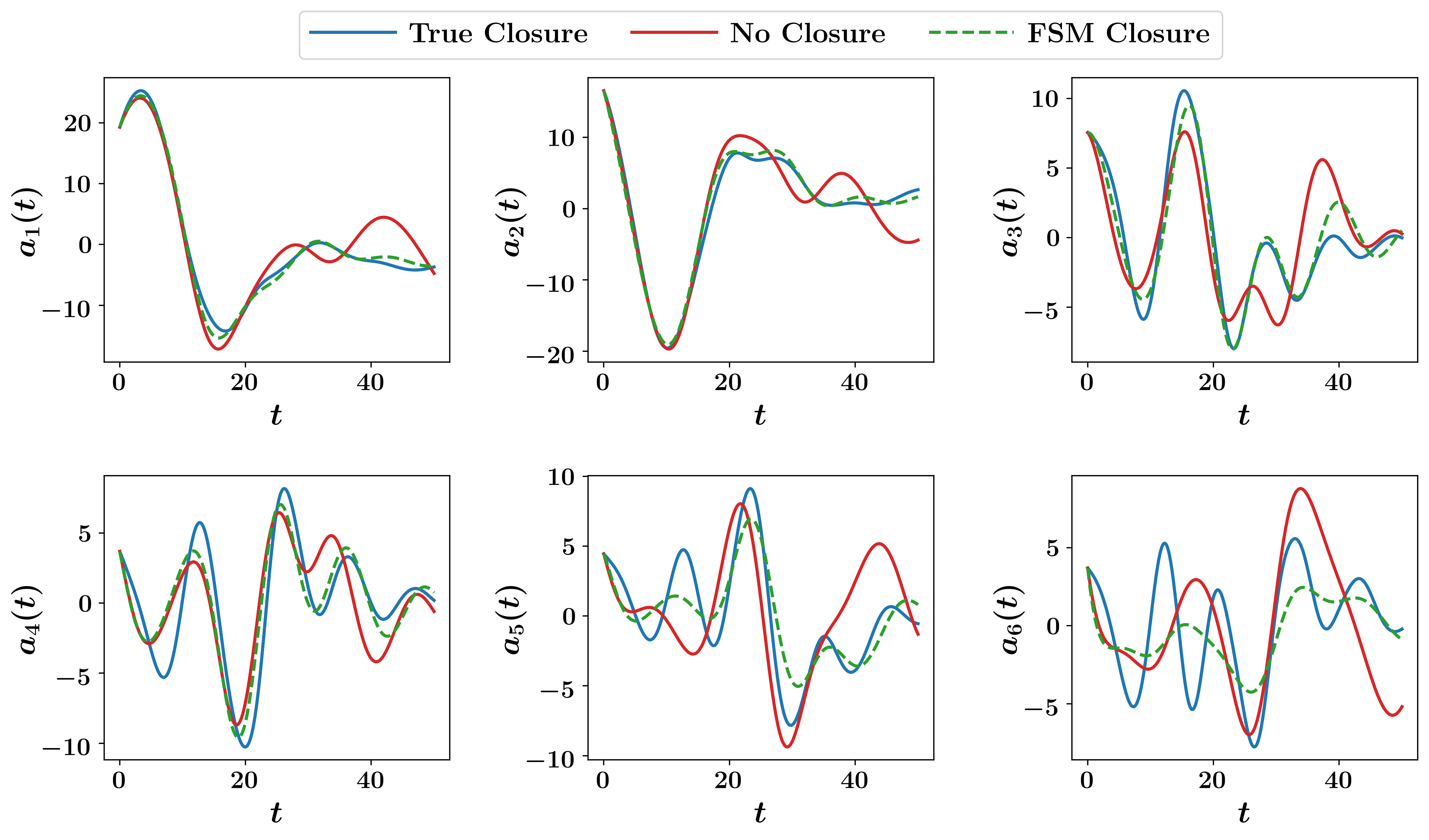}
    \caption{The time evolution of the first 6 modes of the vortex merger problem when full field measurements are collected every $5$ time units.}
    \label{fig:vmcoeff_f}
\end{figure}

We first explore the idealized case where full field measurements of the vorticity fields are collected. We also consider additive Gaussian noise with zero mean and standard deviation of $0.1$. We record data every $5$ time units, corresponding to a total of $10$ measurement instants. We apply the approach presented in \cref{sec:fsm} to compute the mode-dependent parameters $\gamma_i$ and $\beta_i$ for $i=1,2,\dots,n$. The estimated parameters values are then plugged into \cref{eq:closure} to define the closure model. The corresponding predictions of the POD modal coefficients are shown in \cref{fig:vmcoeff_f}, where we see that the higher amplitude oscillations are damped and the ROM trajectory is getting closer to the target values. 

\begin{figure}[ht!]
    \centering
    \includegraphics[width=0.9\linewidth]{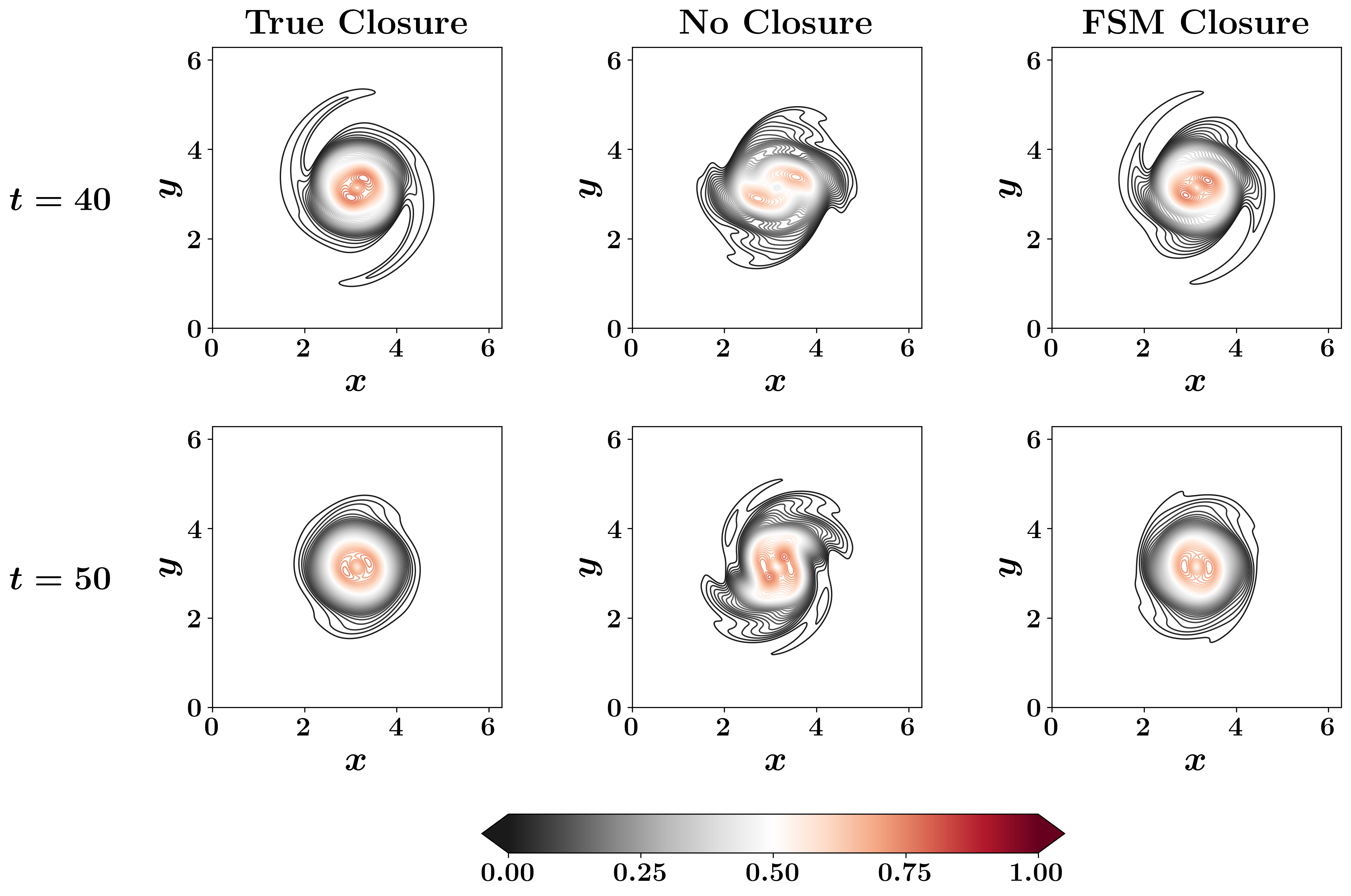}
    \caption{Comparison between the vorticity field at the $t=40$ (top) and $t=50$ (bottom) with True Closure (ground truth from FOM data), No Closure (standard GROM) and the proposed FSM Closure approach with full field measurements.}
    \label{fig:vmfield_f}
\end{figure}

\begin{figure}[ht!]
    \centering
    \includegraphics[width=0.85\linewidth]{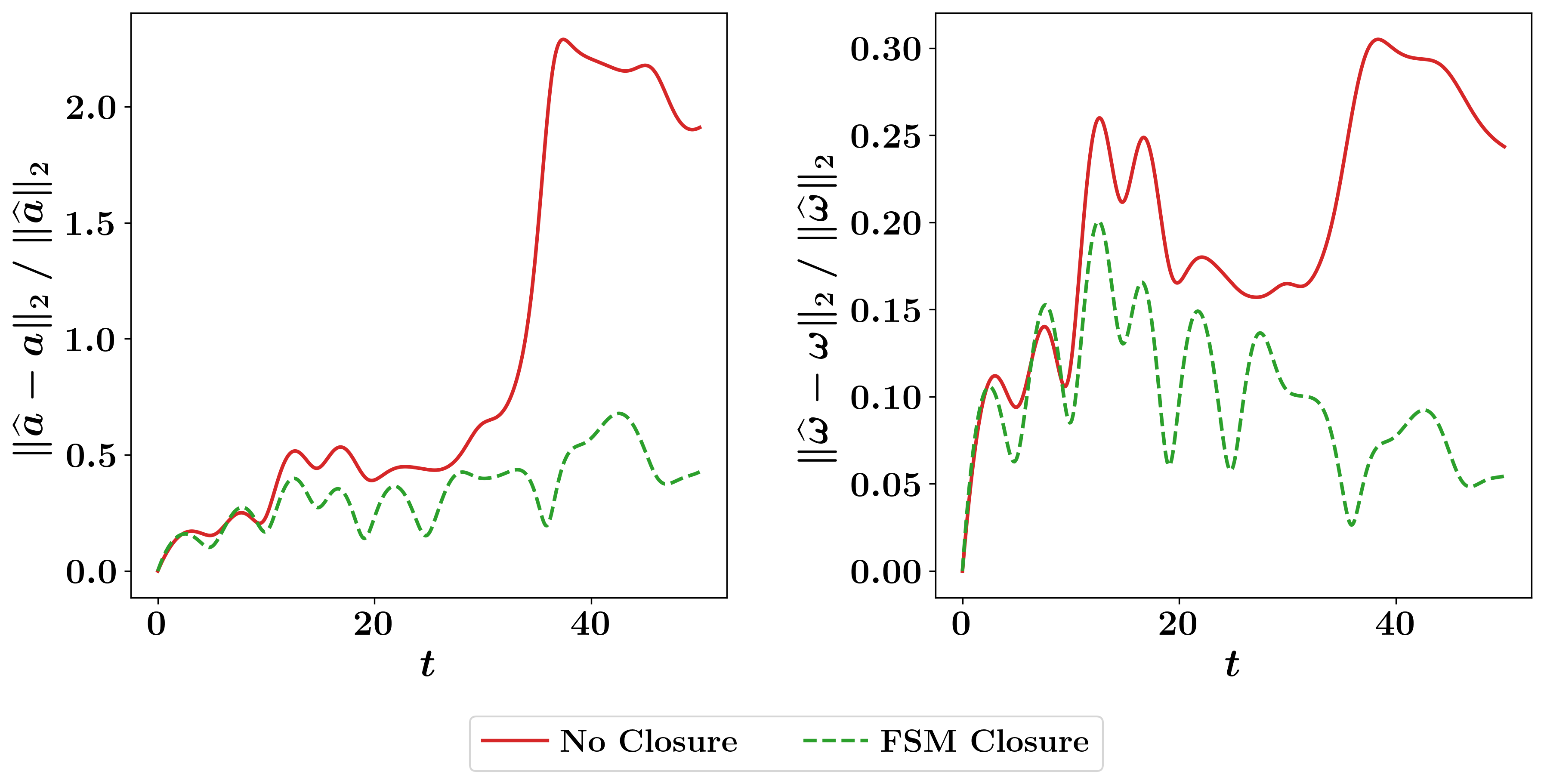}
    \caption{\textcolor{rev}{The relative error between the predicted values for the POD modal coefficients (left) and reconstructed vorticity field (right) compared to their target values for 2D vortex merger problem. Full field measurements are considered for the FSM Closure.}}
    \label{fig:vmerror_f}
\end{figure}

In addition, the reconstruction of the vorticity field at two different time instants is depicted in \cref{fig:vmfield_f}. We see that the GROM model without closure results in flow field predictions that miss significant flow features. On the other hand, the FSM Closure framework is capable of parameterizing the latent control model resulting in a reduced reconstruction error. \textcolor{rev}{\Cref{fig:vmerror_f} also shows the relative error for the predicted POD coefficients as well as the reconstructed vorticity fields as a function of time.}

\subsubsection{Sparse field observations} \label{sub:vmsparse}
\begin{figure}[ht!]
    \centering
    \includegraphics[width=0.9\linewidth]{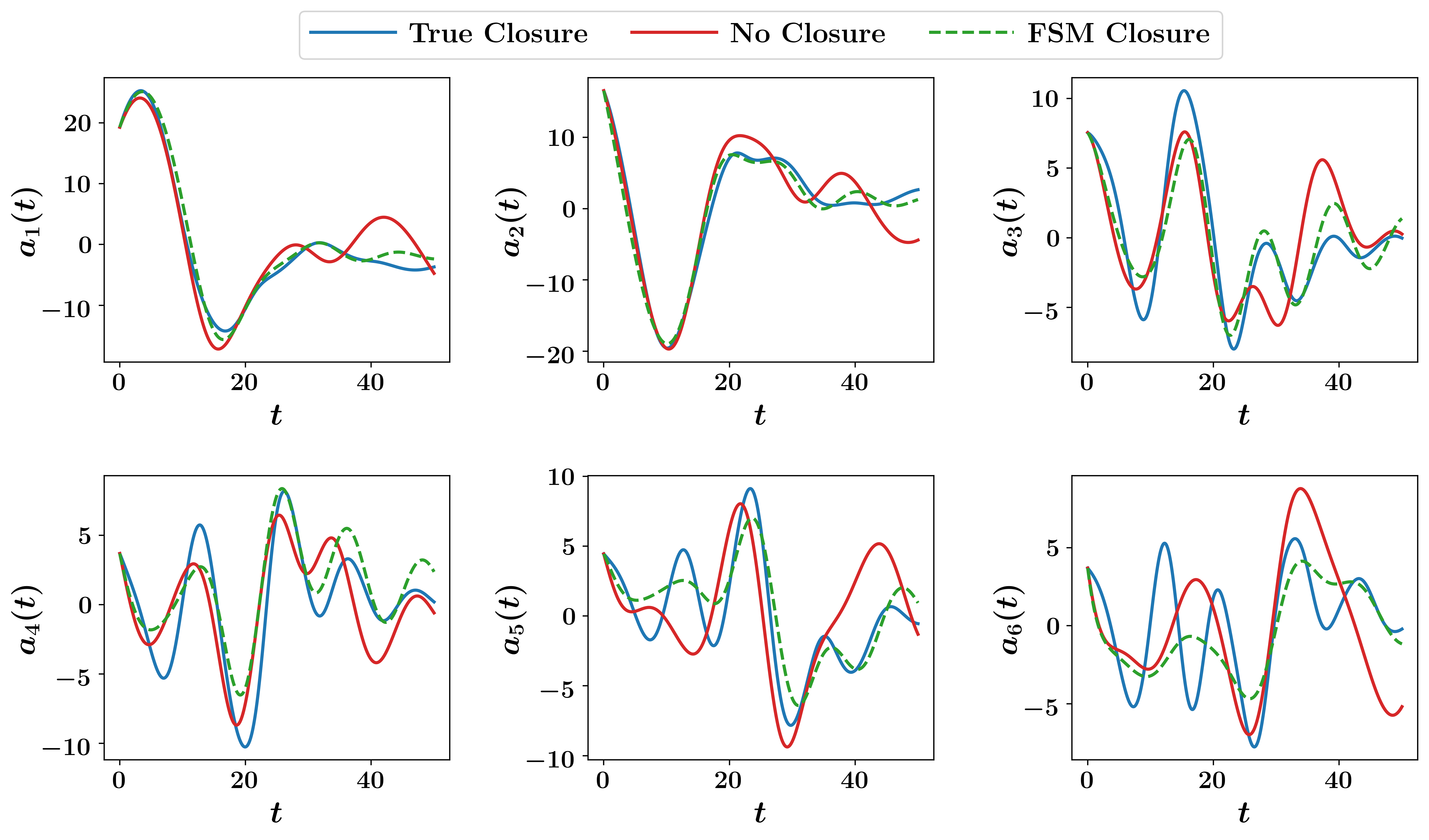}
    \caption{The time evolution of the first 6 modes of the vortex merger problem when only sparse field measurements are available.}
    \label{fig:vmcoeff_s}
\end{figure}

In this section, we investigate the performance of the proposed forward sensitivity approach for mode-dependent control when only spatially sparse observations are available. In particular, we consider a relatively data-scarce regime with $25$ spatial locations (that is less than $0.04\%$ of the total number of grid points). We also incorporate additive Gaussian noise similar to \cref{sub:vmfull}. Although it is typically possible to \emph{clean} this data a bit by considering its spectrum, we intentionally avoid this step to assess the robustness of the FSM framework to data noise and sparsity. We illustrate the predictions of the system's dynamics in the latent space in \cref{fig:vmcoeff_s}. As expected, the predictions of the GROM with FSM closure is quite less accurate than the case with full field measurements (i.e., \cref{fig:vmcoeff_f}) especially at later times. However, compared to the uncontrolled GROM model, we see that the FSM Closure introduces substantial improvements.

\begin{figure}[ht!]
    \centering
    \includegraphics[width=0.9\linewidth]{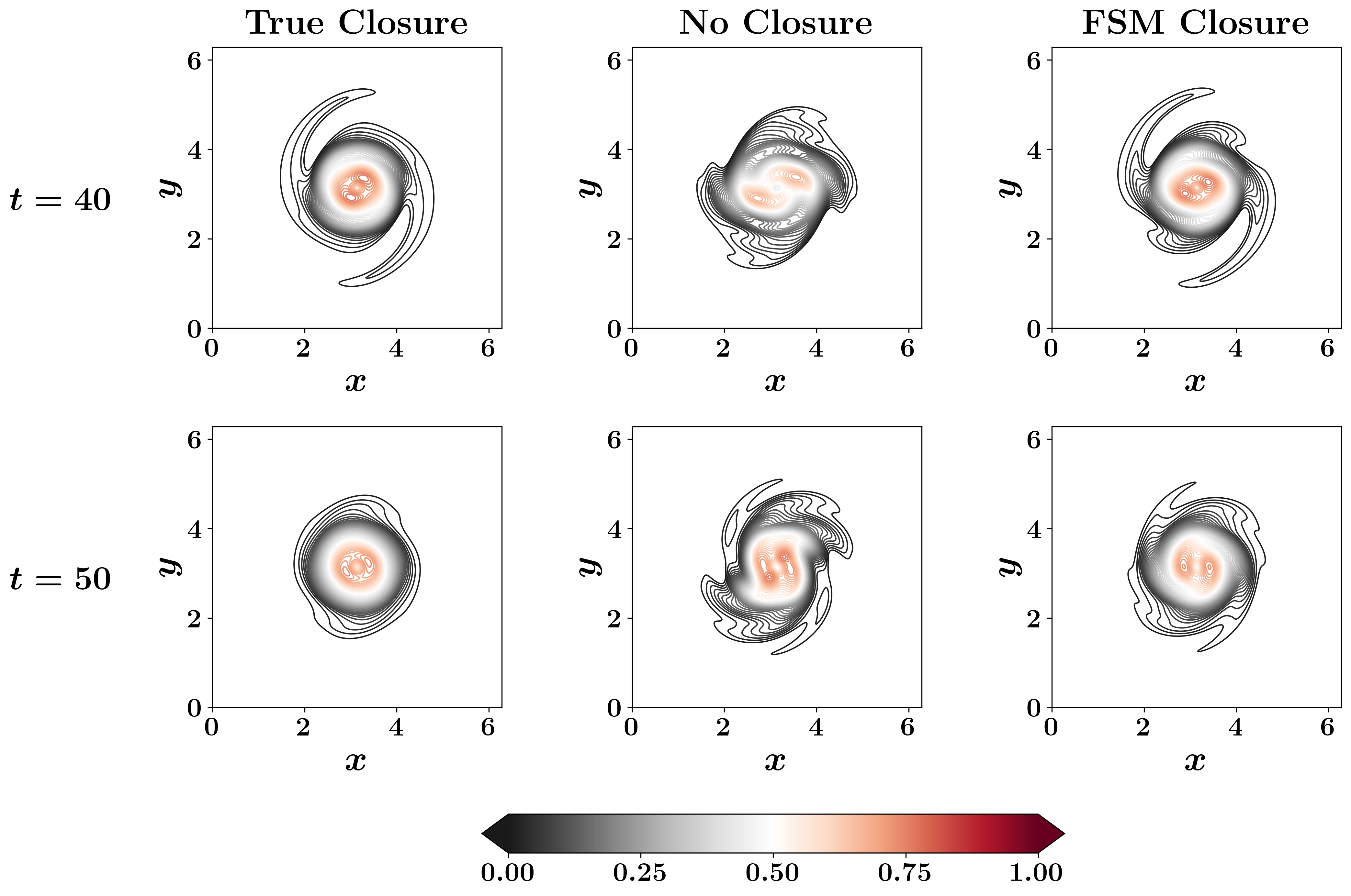}
    \caption{Comparison between the vorticity field at the $t=40$ (top) and $t=50$ (bottom) with True Closure (ground truth from FOM data), No Closure (standard GROM) and the proposed FSM Closure approach with sparse field measurements.}
    \label{fig:vmfield_s}
\end{figure}

\begin{figure}[ht!]
    \centering
    \includegraphics[width=0.85\linewidth]{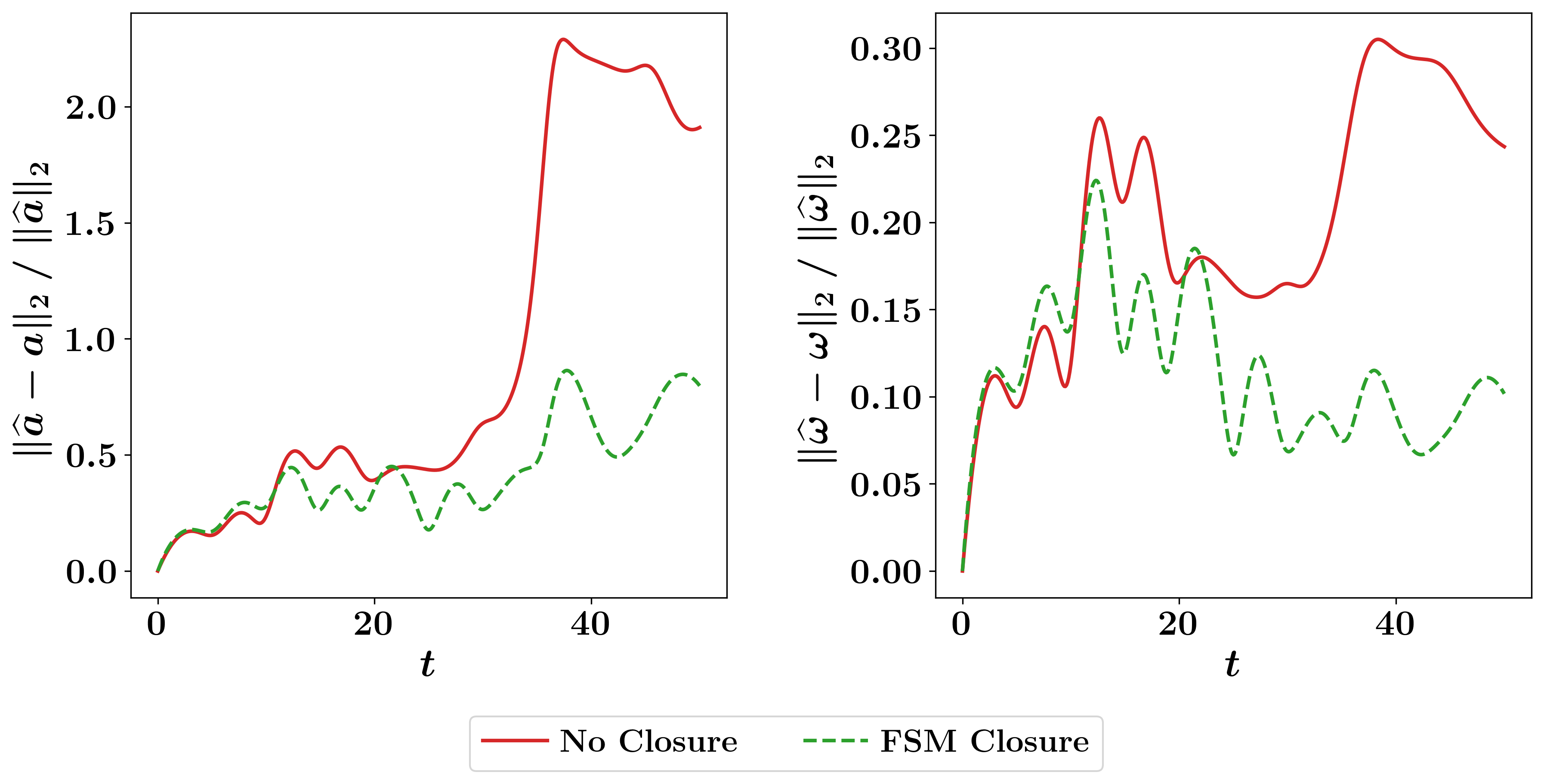}
    \caption{\textcolor{rev}{The relative error between the predicted values for the POD modal coefficients (left) and reconstructed vorticity field (right) compared to their target values for 2D vortex merger problem. Sparse field measurements are considered for the FSM Closure.}}
    \label{fig:vmerror_s}
\end{figure}

\textcolor{rev}{We also observe that the predictions of the first few modes is closer to the target values than the peredictions of the last modes (e.g., $a_5$ and $a_6$). This can be explained by the principle of locality of energy transfer (and modal interactions) of the variational multiscale method \cite{mou2021data,ahmed2022physics}. It implies that the POD truncation results in a model that has much less information about the modes that are closer to the cut-off (e.g., the latest modes) than those that are farther away from the cut-off (e.g., the first modes). In addition, since the first modes have larger contribution to the data construction, the FSM algorithm tends to give higher importance to those mode as it minimizes the error with respect to the measurements. One way to address this issue could be to define a different scaling to ensure that all modal coefficient are equally important.} We also show the reconstructed vorticity fields from the GROM without closure as well as GROM with FSM closure at $t=40$ and $t=50$ in \cref{fig:vmfield_s}. We observe that the FSM closure results in a more accurate recovery of the underlying flow features with respect to the target values (denoted as True Closure). \textcolor{rev}{Finally, \cref{fig:vmerror_s} shows the relative error for the predicted POD coefficients as well as the reconstructed vorticity fields as a function of time.}

\section{Concluding Remarks} \label{sec:conc}
We propose a variational approach for correcting nonlinear reduced order models (ROMs) using the forward sensitivity method (FSM). We cast the closure as a control input in the latent space of the ROM and utilize physical arguments to build parameterized models with damping and dissipation terms. We leverage FSM to blend the predictions from the ROM with available sparse and noisy observations to estimate the unknown model parameters. We apply this approach on a projection based ROM of two test problems with varying complexity corresponding to the one dimensional viscous Burgers equation and the two dimensional vortex-merger problem. These are often considered as canonical test beds for broad transport phenomena governed by nonlinear partial differential equations. We investigate the capability of the approach to approximate optimal values for the mode-dependent parameters without constraining the direction of energy transfer between different modes. Results show that equipping GROM with FSM-based control dramatically increases the ROM accuracy. The predicted trajectories get closer to the values that provide the minimum reconstruction error. The presented framework can effectively enhance digital twin technologies where computationally-light models are required and sensor data are continuously collected.

\section*{Acknowledgments}

The authors are grateful to Sivaramakrishnan Lakshmivarahan for his efforts that greatly helped us in understanding the mechanics of the FSM method. Omer San would like to acknowledge support from the U.S. Department of Energy under the Advanced Scientific Computing Research program (grant DE-SC0019290), the National Science Foundation under the Computational Mathematics program (grant DMS-2012255). 


Disclaimer: This report was prepared as an account of work sponsored by an agency of the United States Government. Neither the United States Government nor any agency thereof, nor any of their employees, makes any warranty, express or implied, or assumes any legal liability or responsibility for the accuracy, completeness, or usefulness of any information, apparatus, product, or process disclosed, or represents that its use would not infringe privately owned rights. Reference herein to any specific commercial product, process, or service by trade name, trademark, manufacturer, or otherwise does not necessarily constitute or imply its endorsement, recommendation, or favoring by the United States Government or any agency thereof. The views and opinions of authors expressed herein do not necessarily state or reflect those of the United States Government or any agency thereof.

\section*{Data Availability}
The synthetic data that support the findings of this study are available within the article. \textcolor{rev}{The complete list of Python scripts that are used in this study can be found in the GitHub page: https://github.com/Shady-Ahmed/fsm-rom-control}.

\bibliographystyle{elsarticle-harv}
\bibliography{references}
\end{document}